\documentclass[pdflatex,sn-mathphys-num]{sn-jnl}

\usepackage{graphicx}%
\usepackage{multirow}%
\usepackage{amsmath,amssymb,amsfonts}%
\usepackage{amsthm}%
\usepackage{mathrsfs}%
\usepackage[title]{appendix}%
\usepackage{xcolor}%
\usepackage{textcomp}%
\usepackage{manyfoot}%
\usepackage{booktabs}%
\usepackage{algorithm}%
\usepackage{algorithmicx}%
\usepackage{algpseudocode}%
\usepackage{listings}%
\usepackage{etoolbox}

\theoremstyle{thmstyleone}%
\theoremstyle{thmstyletwo}%

\theoremstyle{thmstylethree}%

\begin{document}

\title[Article Title]{Pre-training Enables Extraordinary All-optical 
Image Denoising}


\author[1,2]{\fnm{Xudong} \sur{Lv}}\email{lvxudong@hdu.edu.cn}
\equalcont{These authors contributed equally to this work.\\\\\\}

\author[1,3]{\fnm{Yuxiang} \sur{Sun}}\email{yuxiangsun@link.cuhk.edu.cn}
\equalcont{These authors contributed equally to this work.\\\\\\}

\author*[1]{\fnm{Shuo} \sur{Wang}}\email{wangshuo@hit.edu.cn}
\equalcont{These authors contributed equally to this work.\\\\\\}

\author[1]{\fnm{Nanxing} \sur{Chen}}\email{24b321003@stu.hit.edu.cn}

\author[3]{\fnm{Jun} \sur{Guan}}\email{guanjun@cuhk.edu.cn}

\author*[1,4,5]{\fnm{Jingtian} \sur{Hu}}\email{hujingtian@hit.edu.cn}

\affil[1]{\orgdiv{Ministry of Industry and Information Technology Key Lab of Micro-Nano Optoelectronic Information System, Guangdong Provincial Key Laboratory of Semiconductor Optoelectronic Materials and Intelligent Photonic System}, \orgname{Harbin Institute of Technology}, \orgaddress{\city{Shenzhen}, \postcode{518055}, \state{Guangdong}, \country{China}}}

\affil[2]{\orgdiv{School of Electronics and Information Engineering, Zhejiang Provincial Key Laboratory of Intelligent Vehicle Electronics Research}, \orgname{Hangzhou Dianzi University}, \orgaddress{\city{Hangzhou}, \postcode{310018}, \state{Zhejiang}, \country{China}}}

\affil[3]{\orgdiv{School of Science and Engineering}, \orgname{The Chinese University of Hong Kong (Shenzhen)}, \orgaddress{\city{Shenzhen}, \postcode{518172}, \state{Guangdong}, \country{China}}}

\affil[4]{\orgname{Quantum Science Center of Guangdong-Hong Kong-Macao Greater Bay Area}, \orgaddress{\city{Shenzhen}, \postcode{518055}, \state{Guangdong}, \country{China}}}

\affil[5]{\orgdiv{Key Laboratory of Photonic Technology for Integrated Sensing and Communication, Ministry of Education}, \orgname{Guangdong University of Technology}, \orgaddress{\city{Guangzhou}, \postcode{200124}, \state{Guangdong}, \country{China}}}

\abstract{Optical neural networks are emerging as powerful machine learning and information processing tools because of their potential advantages in speed and energy efficiency. The training methods of these physical models, however, remain underexplored compared to their digital counterparts and are leading to suboptimal performance. This paper reports a pre-training-driven approach that leads to snapshot image denoising with substantially improved quality. We demonstrated effective free-space optical denoising by a diffractive network optimized by a two-step process including (1) pre-training using a massive dataset of 3.45 million diverse but simple images and (2) fine-tuning with the corresponding task-specific datasets. Compared to conventional Fourier-domain filtering and directly trained diffractive networks, such a transfer learning process exhibited prominent advantages for denoising images degraded by severe noise, peak signal-to-noise ratio (PSNR) below 8 dB, while preserving fine image features and improving the PSNR to above 18 dB. Importantly, the same pre-trained optical network could be consistently fine-tuned to process degraded images from highly diverse styles ranging from handwritten digits (MNIST) and chest X-rays (ChestMNIST) to CIFAR-10 images and human faces (CelebA). We further demonstrated the critical role of our optical denoisers in vision-based applications, including face detection, plate recognition, and localization of UAVs in noisy conditions.}




\maketitle

\section{Introduction}\label{sec1}
Artificial neural networks (ANNs) have become powerful tools for visual information processing in technologies including image denoising \cite{gondara2016medical, devalla2019deep, nienhaus2023live}, face detection \cite{qi2022yolo5face}, and video-based surveillance \cite{yu2024yolo}. The processing speed and energy consumption of ANNs, however, can become the limiting factor for deployment in edge devices equipped with restricted computing power \cite{floreano2015science, zhou2022swarm, xu2025intelligent, wu2025intelligent}. Optical neural networks are emerging as the next-generation solution for artificial intelligence because of their advantages in processing speed and energy efficiency \cite{liu2021research, fu2024optical, shen2017deep, hua2025integrated, ahmed2025universal,kalinin2025analog, fu2023photonic, liao2023integrated}. In particular, diffractive networks have demonstrated the ability to perform all-optical machine learning \cite{lin2018all, zhou2021large} and visual data processing \cite{luo2022computational, icsil2024all, hu2024subwavelength} in a fast, single-shot manner, i.e., within a single pass of free-space light propagation \cite{guo2026deeper}. Still, the applications of these free-space processors are often restricted to relatively simple image datasets such as MNIST \cite{yu2025all} and Fashion-MNIST \cite{huang2026anti, he2023pluggable, chen2023all} while achieving limited success on more challenging tasks and visual content \cite{xue2024fully, huo2023optical}. Such issues in performance and versatility are often attributed to the lack of nonlinearity that constrains the learning capabilities of these optical models \cite{li2024nonlinear, yildirim2024nonlinear, zhou2025all, chen2023deep, ning2025multilayer}. In contrast, the training of optical neural networks, although receiving increasing attention \cite{wetzstein2020inference, wright2022deep, momeni2023backpropagation}, remains insufficiently investigated, but also critical for the optimization of deep learning models. Most existing free-space diffractive networks still rely on backpropagation-based in-silico training in a single-step, end-to-end manner \cite{zangeneh2021analogue, momeni2025training, hu2024diffractive}, which can result in suboptimal convergence and limited generalization capability.

Pre-training has been a critical strategy for developing ANNs because of its ability to promote faster convergence towards improved performance and robustness \cite{he2019rethinking, zoph2020rethinking}. Compared to training a model from scratch (i.e., randomly initialized weights) for a specific task, a model that has already been trained on a large, generic dataset provides a more appropriate initialization with better prior understanding of the fundamental features \cite{hendrycks2019using}, where the learned knowledge can be effectively transferred to the downstream tasks through a quick fine-tuning process \cite{chen2021pre}. However, this method has rarely been applied to the training of optical neural networks, possibly because a suitable dataset is lacking. In particular, pre-training of vision tasks often adopts the ImageNet dataset \cite{deng2009imagenet}, which is overly complex for diffractive networks, where learning relevant features remains highly challenging. Therefore, the adoption of scaled-up but simple datasets with highly diverse data is crucial for advancing diffractive processors with pre-training techniques towards robust performance on real-world applications.

\begin{figure}[thpb]
    \centering
    \includegraphics[width=0.95\linewidth]{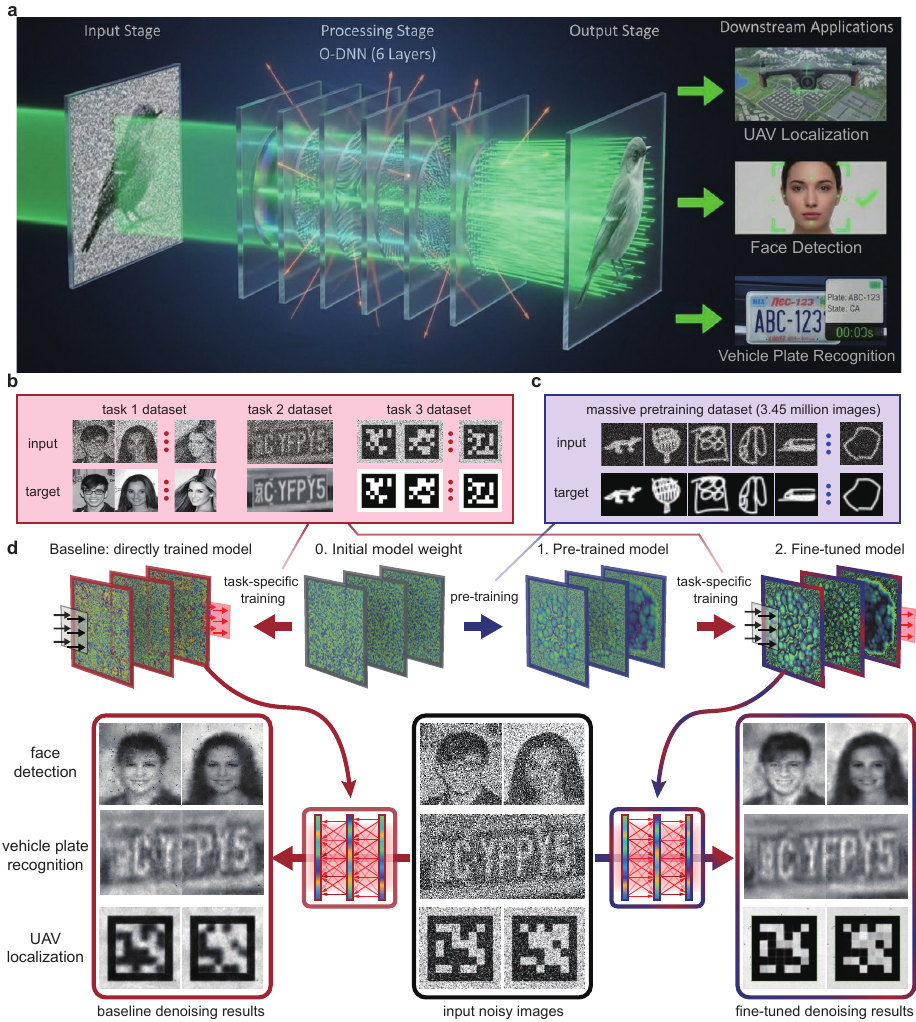}
    \caption{\textbf{Pre-training-based optimization framework for high-performance snapshot diffractive denoising in free-space.} (a) Schemes depicting the overall architecture of our diffractive network for all-optical image denoising. Representative input and ground-truth images from (b) task-specific datasets used for direct training or fine-tuning and (c) a 3.45-million-image dataset for pretraining the optical model. (d) Side-by-side comparison between the procedure of conventional direct training (train-from-scratch) from randomly initialized weights and fine-tuning based on a pre-trained model. The same pre-trained model was used for the task-specific fine-tuning of all datasets.}
    \label{F1}
\end{figure}

Here, we showed a pre-training approach for the development of free-space optical processors that realized snapshot image denoising with unprecedented accuracy and fidelity (Figure \ref{F1}). For this pipeline, we constructed a massive pre-training dataset with 3.45 million pairs of clean and noisy images based on a fraction of QuickDraw \cite{ha2018neural}, which consisted of simple but highly diverse binary images. We demonstrated that the diffractive network pre-trained on this dataset could be universally fine-tuned to denoise broad categories of images, ranging from handwritten digits and fashion products (i.e., EMNIST \cite{cohen2017emnist} and Fashion-MNIST \cite{xiao2017fashion}) to human faces (i.e., CelebA \cite{zhang2020celeba}) and vehicle plates (i.e., CBLPRD \cite{cblprd330k}). Remarkably, this knowledge-transfer strategy enabled effective denoising of heavily degraded images with a peak signal-to-noise ratio (PSNR) $<8$ dB to produce nearly clean images with PSNR $>18$ dB, which was impossible for the diffractive networks trained from scratch on the corresponding datasets. The model also showed superb generalizability towards different types of degradations, including Gaussian, Poisson, and salt \& pepper noises, as well as their combinations. We experimentally verified the proposed diffractive network in a reflective configuration for performing denoising on amplitude-encoded image information. The system, built with a phase-only SLM and mirror-assisted multiple diffraction, successfully executed image denoising in real time, with a processing speed limited only by the light source intensity and image sensor sensitivity.

The training framework we established improved the denoising capabilities of optical models substantially so that the system exhibited tremendous potential for practical applications based on vision systems. We first targeted a scenario where the captured images of human faces were severely degraded by simulated environmental noises that prohibited face detection by RetinaFace \cite{deng2020retinaface}. Our diffractive processor, fine-tuned from the pre-trained model, demonstrated the ability to recover the facial features from noisy celebrity images to enable effective face recognition, which was not possible with either Fourier-domain filtering or the trained-from-scratch diffractive networks. We further showed the application of our diffractive denoiser in assisting accurate recognition of vehicle plates \cite{ke2023ultra} and the localization of unmanned aerial vehicles (UAVs) \cite{olson2011apriltag}. Overall, we believe that our pre-training framework represents a promising strategy that can significantly advance the development of high-performance optical networks. The single-shot, all-optical image denoiser optimized by this method also significantly outperforms existing optical denoising systems, paving the way for numerous visual information tasks in noisy environments.

\section{Results and Discussion}\label{sec2}
\subsection{Training of diffractive denoisers by knowledge transfer}\label{sec2_subsec1}
We demonstrated the capability of our pre-training and transfer-learning pipeline for diffractive networks through the task of all-optical image denoising, which has a broad impact on vision-based technologies, including plate recognition, face detection, and UAV localization. Figure \ref{F1}a depicted the architecture of the free-space processor that performs denoising by successive diffractive layers. For all numerical demonstrations, the model consisted of 6 diffractive layers with $270\times 270$ units (each with a size of $0.47\lambda\times0.47\lambda$). The degraded images were represented by the amplitude-encoded optical wavefronts at the input plane and processed by a combination of light propagation and phase modulation by the diffractive layers (see Methods for details), forming denoised images at the detector plane. Figure \ref{F2}b showed the task-specific datasets we aim to process. Separate optical models were trained to reconstruct from noisy CelebA, CBLPRD, and AprilTag \cite{olson2011apriltag} images to facilitate face detection, plate recognition, and UAV localization applications, respectively. Performance on other benchmark datasets such as EMNIST, FashionMNIST, CIFAR-10 \cite{cifar10_dataset}, and MedMNIST \cite{yang2023medmnist} was also quantified to evaluate the versatility and generalizability of our method. For all datasets, the input to the optical models was prepared by adding Gaussian noise (with noise probability $\gamma=0.5$) to the clean images, which were also used as the ground truth for the expected output. Figure \ref{F1}c further presents the pre-training dataset constructed based on the QuickDraw dataset, where the images before and after noise addition are used as the target and input of the models. We selected 3.45 million images where 10,000 images were chosen from each of the 345 QuickDraw categories. (see Methods for the preparation of datasets). Although the dataset consisted of only simple binary drawings, its massive size and diversity promote the learning of robust, generalizable features for effective denoising by the pre-trained diffractive network.

Figure \ref{F1}d further depicted the overall procedure of our training method based on pre-training and transfer learning in a side-by-side comparison with the conventional training-from-scratch approach (baseline model). In the latter case, diffractive denoisers for each type of image inputs were directly trained by the corresponding datasets, starting from randomly initialized weights. For all datasets, the denoised images produced by the directly trained model (baseline model) suffered from prominent imperfections, including artifacts, distortions, and blurriness. We addressed these issues by first training the randomly initialized diffractive network using the massive QuickDraw-based dataset to obtain a sufficiently pre-trained optical model, which learned the relatively general-purpose features relevant for image denoising. Then, the pre-trained model was fine-tuned using the task-specific datasets to achieve high-fidelity image denoising in the respective applications (see Methods and Supplementary Sect. 1 for details of training). The transfer-learning process allowed the diffractive networks to perform image denoising with unprecedented quality, producing well-defined features and morphologies consistent with the ground truth.

\subsection{Benchmark test of diffractive denoisers by pre-training}\label{sec2_subsec2}
\begin{figure}[thpb]
    \centering
    \includegraphics[width=\linewidth]{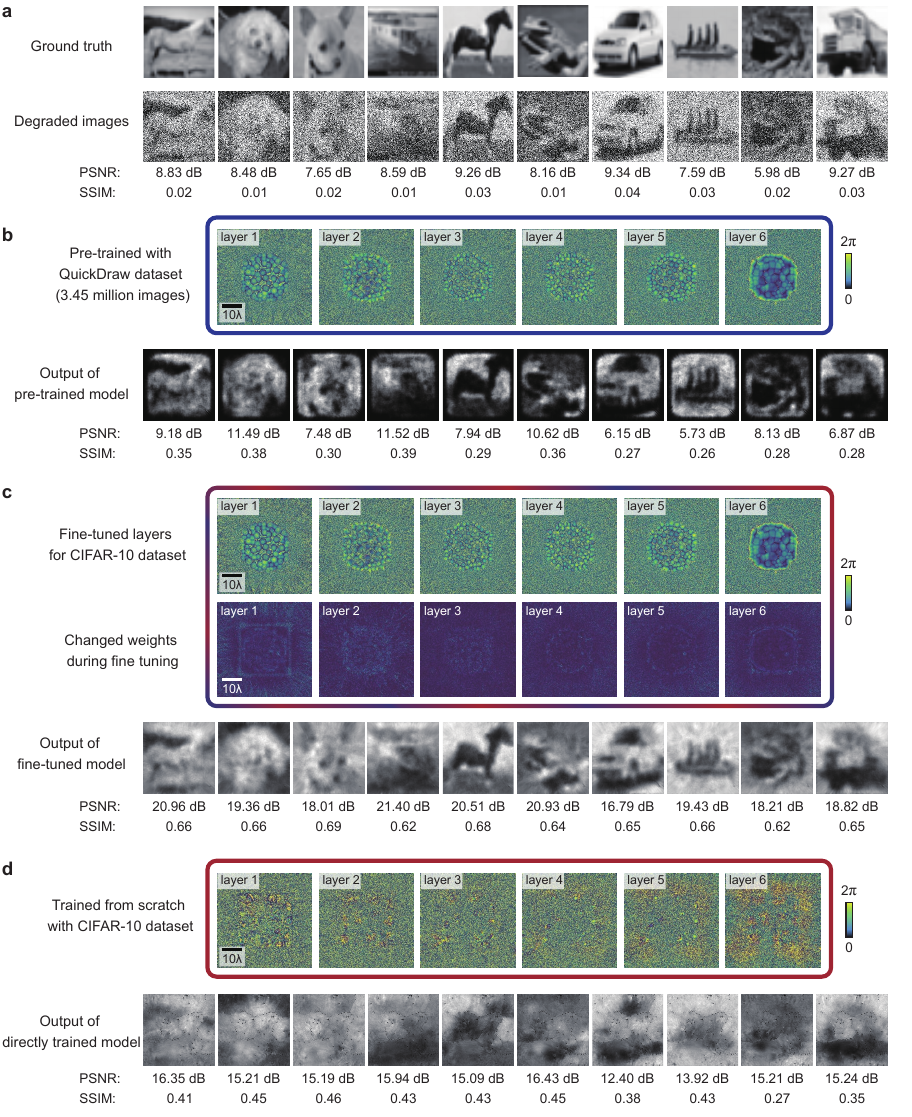}
    \caption{\textbf{Numerical benchmark tests for diffractive denoiser obtained by knowledge transfer from a pre-trained model.} (a) Representative ground-truth and noisy input images from the test dataset based on CIFAR-10. Design and all-optical denoising output for diffractive networks optimized through (b) only pre-training (using the QuickDraw-based dataset), (c) pre-training and fine-tuning with CIFAR-10, and (d) direct training with CIFAR-10 (i.e., the trained-from-scratch baseline model).}
    \label{F2}
\end{figure}

We numerically tested the performance of our diffractive processors on denoising images with broad categories of visual content to demonstrate the advantage and versatility of the pre-training and knowledge transfer approach. Figure \ref{F2} showed the representative benchmark results for denoising CIFAR-10 in comparison with the baseline model. The results and analysis for other datasets (including ChestMNIST, BloodMNIST, CelebA, CBLPRD, and AprilTag) could be found in Supplementary Sect. 2. Figure \ref{F2}a illustrated a few pairs of ground-truth and input images from the grayscale CIFAR-10 dataset. With the addition of Gaussian noise ($\gamma=0.5$), the PSNR of the images was degraded to $\sim 8$ dB, and their structural similarity index measure (SSIM) compared to the ground truth was reduced to $<0.04$. Figure \ref{F2}b showed the phase-modulation layers of the pre-trained diffractive network (using QuickDraw images) and their test results for denoising CIFAR-10 images, where the entire dataset and style have never been presented to the model. Without fine-tuning, the model already showed the ability to recover a significant amount of the original features but still suffered from compromised image quality (SSIM $\sim 0.30$, PSNR $\sim 10$). Figure \ref{F2}c further presented the design of the network after fine-tuning with the CIFAR-10 dataset and the associated denoising outputs. Notably, the phase-modulation distributions of the diffractive layers after the fine tuning closely resembled the pre-trained model (with only subtle differences as highlighted), but achieved significantly improved denoising outcomes (SSIM $\sim 0.65$, PSNR $\sim 20$). Importantly, this network successfully recovered features of the original images, including the appearance of vehicles and animals, which were not possible by the pre-trained (Figure \ref{F2}b) and baseline (Figure \ref{F2}d) models. The superiority of the pre-training \& fine-tuning approach was consistently verified for all other categories of image denoising tasks (Supplementary Sect. 2). Comparison between the diffractive layers further showed that the baseline network was drastically different from the fine-tuned model, with significantly better denoising capabilities. Such a difference likely arose from the trapping of the backpropagation algorithm within suboptimal local minima during direct training or the overfitting of the model when trained with limited data.

Through Fourier space analysis, we further elucidated the mechanistic origin for the superb performance of our pre-trained model (Supplementary Figure S9). Conventional Fourier-space filtering could achieve optical denoising in a 4-f system by removing high-frequency spatial modes corresponding to noise information (Supplementary Figure S9-b), but inevitably caused the loss of fine features from the images and artificial background textures. The baseline model (trained from scratch) was able to retain certain high-frequency features, while also kept severe noise modes in the background. The pre-training strategy, in contrast, could simultaneously achieve effective removal of noise modes while retaining the key fine features of the clean images (Supplementary Figures S9c). Quantitative analysis (Supplementary Figures S9d) over three test datasets further showed that the pre-training strategy outperformed the 4-f system and baseline model by achieving the best evaluation metrics (SSIM $\sim0.66$, PSNR $\sim20$). Importantly, the design approach of our free-space denoiser was scalable for handling larger images for potential high-throughput image processing (Supplementary Figure S7 and S8). Although the entire training process was based on Gaussian noises ($\gamma=0.5$), the optimized model demonstrated superb generalization for handling other types of noises, such as Poisson and salt \& pepper, as well as their mixtures (Supplementary Figure S10).

We analyzed the relationship between the architectural depth of the diffractive neural network and its denoising performance across multiple datasets. As shown in the weight visualizations (Supplementary Figure S11a), increasing the layer count from 1 to 6 leaded to significantly more intricate and granular phase patterns, transitioning from nearly uniform distributions to complex, high-frequency spatial features that enable sophisticated hierarchical light-field modulation. This evolution in feature representation directly translated to a consistent, monotonic improvement in quantitative metrics (Supplementary Figure S11-b), where both PSNR and SSIM exhibited a sharp leap between 2 and 4 layers. This trend confirmed that increased optical depth is essential for effectively decoupling noise from the underlying signal, consistently enhancing the model's capacity to restore structural integrity regardless of the dataset's complexity. We further systematically evaluated the impact of other physical structural parameters on the denoising performance of the diffractive neural network. As illustrated in (Supplementary Figure S12a and S12b), increasing the axial distance between diffractive layers does not lead to a significant monotonic change in PSNR or SSIM across the four datasets. It suggested that the network’s diffraction-based feature extraction is robust to longitudinal scaling within the tested range. Similarly, as shown in Supplementary Figure S12c and S12d, increasing the size of the diffractive layers (i.e., the number of neurons per layer) beyond an initial threshold yields diminishing returns, with both PSNR and SSIM metrics reaching a stable plateau. This indicated that while a minimum spatial complexity was required to capture essential signal features, further expansion of the layer size provided marginal gains in denoising fidelity, highlighting the model's efficiency in maintaining high structural similarity and signal-to-noise ratios even with compact physical footprints.

\subsection{Experimental demonstration of pre-training-enabled optical denoiser}\label{sec2_subsec3}
\begin{figure}[thpb]
    \centering
    \includegraphics[width=0.95\linewidth]{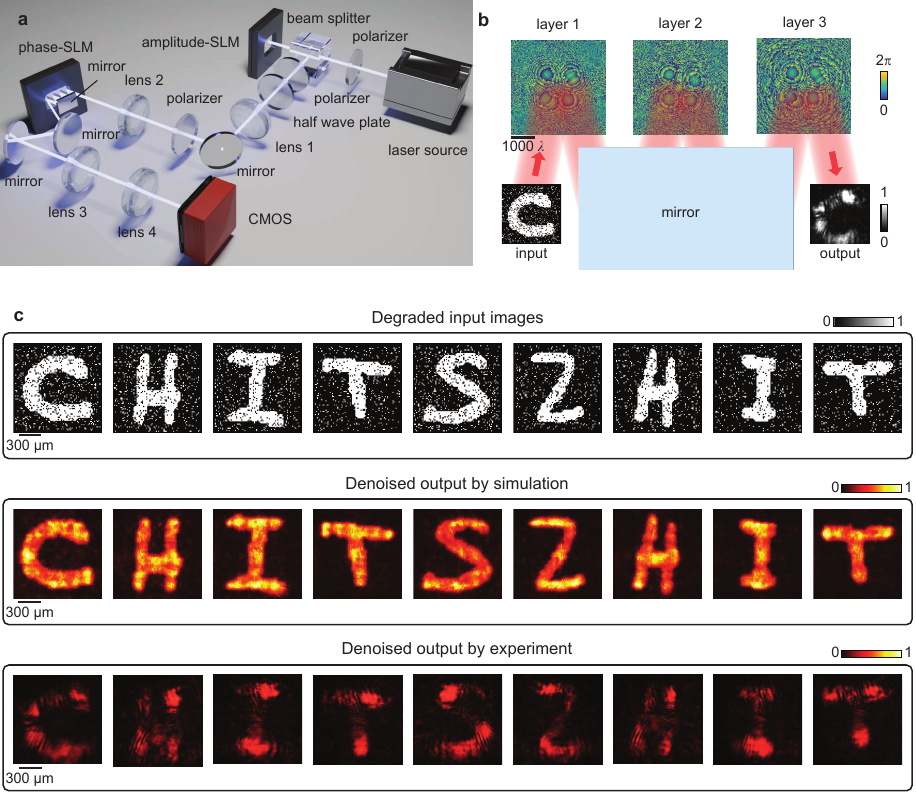}
    \caption{\textbf{Experimental demonstration of the pre-training-enabled all-optical image denoiser.} (a) Schematic diagram of the experimental system for validating the pre-training-enabled optical denoiser. A supercontinuum laser source, spectrally filtered to 635 nm, illuminated an amplitude-only SLM to encode degraded input images. The reflected beam was relayed by a 4-f system into a reflective cavity-type diffractive processor consisting of a phase-only SLM and a mirror, and the denoised output was finally recorded by a CMOS camera through the imaging optics. (b) Enlarged schematic of the reflective diffractive processor. Three diffractive layers were laterally multiplexed on the phase-only SLM, and the optical field was sequentially modulated through mirror-assisted multiple reflections before forming the output image. The phase distributions of the three layers are shown above, with phase values spanning 0 to $2\pi$. (c) Representative denoising results for EMNIST samples not used during training. The top row shows degraded input images, the middle row shows the corresponding denoised outputs predicted by simulation, and the bottom row shows the experimentally recorded outputs. Despite experimental non-idealities, the reconstructed images preserved the main character morphologies and exhibited clear suppression of background noise, supporting the physical feasibility of the pre-training-enabled diffractive denoiser.}
    \label{F3}
\end{figure}

To verify that the performance gain enabled by pre-training could be translated to a physical platform, we experimentally implemented the optical denoiser in a reflective configuration Figure \ref{F3}a, b. A supercontinuum laser source (SC-5, NKT Photonics), spectrally filtered to a central wavelength of 635 nm, was used as the illumination source. The degraded input image was first encoded on an amplitude-only spatial light modulator (SLM) and, after reflection, relayed through a 4-f beam-reduction system to a cavity-type diffractive processor consisting of a phase-only SLM (RSL, China) and a planar mirror. The phase SLM and the mirror were separated by 4 cm, which reduced the system footprint while improving pixel utilization. To relax alignment tolerance in the experiment, each diffractive neuron was mapped onto a 2 × 2 block of SLM pixels with a pitch of 3.6 $\mu$m, corresponding to an effective neuron size of 7.2 $\mu$m. Three diffractive layers, each containing 350 × 350 neurons, were laterally multiplexed on the same phase SLM. The incident beam was directed onto the first layer at a designed off-axis angle, then successively routed to the second and third layers through mirror-assisted reflections, before forming the denoised output after free-space propagation. The input and output fields were both defined over a 1.15-mm field of view. To accurately describe the tilted propagation trajectory in this reflective architecture, the inter-layer transport was modeled using the off-axis angular spectrum method, as detailed in Methods. 

Figure \ref{F3}c presented representative experimental results for EMNIST samples that were not used during training. Noise was added to the input characters, degrading the images to a PSNR of $\sim 2$ dB and an SSIM of $\sim 0.20$. The resulting corrupted wavefronts were then processed by the diffractive system in a single optical inference cycle, and the reconstructed outputs were recorded by a CMOS camera through the imaging optics. Despite the inevitable non-idealities associated with physical implementation, the experimental outputs preserved the main character morphologies and exhibited clear suppression of background noise, in good agreement with the numerical predictions. These results confirmed that the pre-training-enabled diffractive network retains its denoising functionality after deployment in hardware, and support its use as a compact optical front-end for subsequent machine-vision tasks.

\subsection{Pre-training-enabled all-optical denoising for downstream vision tasks}\label{sec2_subsec4}
\begin{figure}[thpb]
    \centering
    \includegraphics[width=\linewidth]{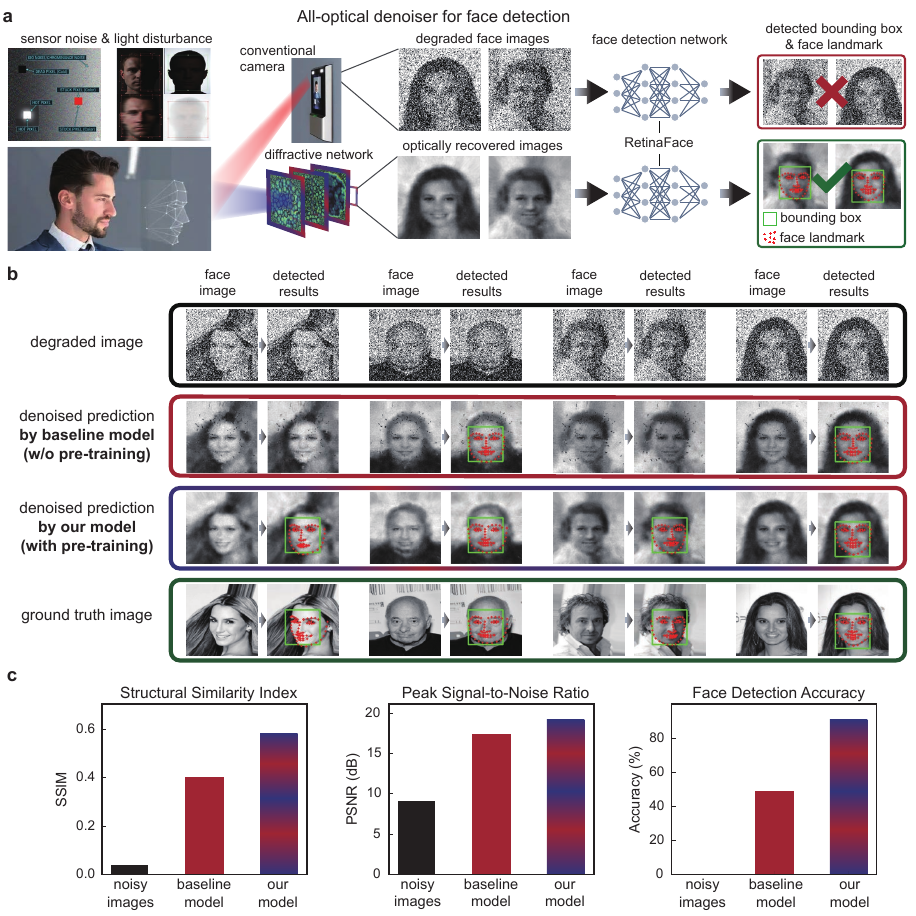}
    \caption{\textbf{Pre-training-enabled snapshot all-optical denoisers for face detection.} (a) Preprocessing of noisy images by pre-sensor diffractive networks optimized by pre-training and transfer learning for effective face detection. (b) Face detection by RetinaFace for denoised outputs recovered from corrupted images recovered by our diffractive processor in a side-by-side comparison with the trained-from-scratch baseline model. (c) Quantitative performance evaluation of the denoising performance for the CelebA dataset using metrics including SSIM, PSNR, and face-detection accuracy. The statistics were collected over 1400 test images with varied gender, age, hair, and facial expression.}
    \label{F4}
\end{figure}

The pre-training-based framework has significantly improved the performance of diffractive networks so that they can be used for image processing tasks in real-world applications. Figure \ref{F4}a showed a proof-of-concept illustration of noise-resilient face detection systems based on the established optical processors. Existing face detection systems could fail to produce accurate results when the captured images were corrupted by environmental and sensor noises. To address this issue, our diffractive network achieved effective pre-sensor denoising that preprocessed the images for detection by the RetinaFace model. Figure \ref{F4}b showed the face detection results for images with noise ($\gamma=0.5$) and the denoised output by our optical model in comparison to the baseline (trained from scratch). Before being processed by the diffractive network, the corrupted images of celebrity faces exhibited poor PSNR and SSIM compared to the original images, thus prohibiting any effective face detection. Processing by the baseline model trained directly using the CelebA dataset (see Methods and Supplementary Figure S1 for the details of training) improved the success rate of face detection, but still failed on a considerable fraction of images. Instead, the diffractive network optimized by pre-training and fine-tuning could output a denoised image with sufficiently high fidelity to allow robust face detection at a high success rate. Compared to the baseline model, which exhibited significant artifacts and geometric blurring, our model achieved higher detection accuracy and produced face landmarks (red points) and bounding boxes (green squares) that are more closely aligned with the ground truth. Specifically, the facial features restored by our method maintained high structural consistency with the original clean images, effectively preserving the anatomical integrity required for precise landmark localization. Figure \ref{F4}c further quantified the performances over 1439 randomly selected facial images, which have not been used for the training/fine-tuning. Overall, the pre-training strategy significantly improved the PSNR ($\sim 19.20$ dB) and SSIM ($\sim 0.59$) of the processed images with a substantial advantage over the conventional approach (PSNR $\sim 17.32$ dB, SSIM $\sim 0.40$). The additional fidelity allowed RetinaFace to perform face detection at a significantly higher accuracy ($91.33\%$ versus $49.50\%$).

The comparative analysis of the diffractive layer weights and the corresponding denoising performance on the CelebA dataset highlighted the efficacy of our transfer learning strategy (see Supplementary Figure S4). As illustrated in the Supplementary Figure S4b, the pre-trained model (trained on QuickDraw) exhibited a foundational ability to suppress global noise, yet its phase masks lacked the high-frequency structural information required to resolve intricate facial features. The baseline model (Supplementary Figure S4d), trained from scratch on the limited CelebA subset, suffered from overfitting and noise-induced blurring, resulting in distorted reconstructions where anatomical landmarks are indistinguishable. In contrast, the fine-tuned model (Supplementary Figure S4c) achieved superior high-fidelity reconstruction by optimizing its phase modulation through the transfer learning. The resulting phase masks demonstrated more complex, granular spatial distributions that successfully recovered sharp contours (e.g. eyes, nose, and mouth). This transition from “fail” to “success” underscored the critical role of pre-training-enabled knowledge transfer in adapting diffractive processors to high-dimensional, real-world image domains. The tremendous performance boost made free-space diffractive computing a promising technology for low-latency face detection systems that are robust towards noise.

\begin{figure}[thpb]
    \centering
    \includegraphics[width=0.99\linewidth]{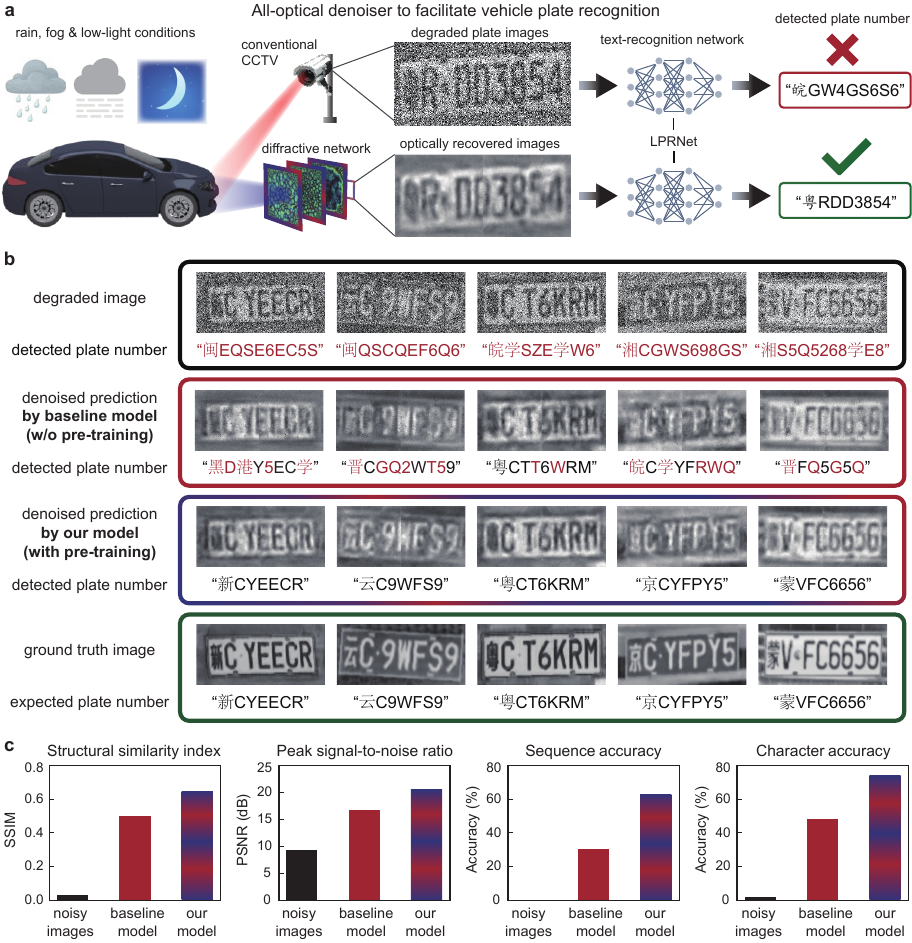}
    \caption{\textbf{Pre-training-enabled snapshot all-optical denoisers for vehicle plate recognition.} (a) Schematic of the pre-sensor diffractive network, which utilizes pre-training and transfer learning frameworks to optimize optical preprocessing for robust vehicle plate recognition. (b) The comparative evaluation results showcased denoised outputs from our diffractive processor against the baseline model (trained from scratch), demonstrating superior restoration of corrupted images. (c) Quantitative assessment of denoising and recognition performance on the CBLPRD dataset. Metrics include SSIM, PSNR, sequence accuracy, and character accuracy, with statistics aggregated over 1100 test images featuring diverse backgrounds and character configurations.}
    \label{F5}
\end{figure}

Figure \ref{F5}a presented the application of our optical denoising model for vehicle plate recognition with strong environmental noise. For this task, we showcased the capabilities of our pre-training-enabled optical model for denoising images from the CBLPRD dataset, where the plate consisted of a combination of numbers, letters, and Chinese characters. Due to the aspect ratio of the plate images (2:1) differed significantly from the pre-training dataset (1:1), they were split into two square-shaped images and processed separately by the diffractive denoisers. Figure \ref{F5}b further compared the quality of image denoising by the pre-training model in comparison to the baseline approach for this recognition task. The knowledge transfer from the pre-trained network allowed our denoiser to recover the letters and digits with clear, well-defined morphology while suppressing the background noises and artifacts, thereby facilitating successful text recognition. Although the fine strokes of the Chinese characters were not always resolvable after denoising, the reconstructed contour still allowed successful identification of the characters. In contrast, the images processed by the trained-from-scratch model still exhibited severe degradation by noises and distortions that prevented accurate detection of plate numbers. Figure \ref{F5}c provided a quantitative analysis of the denoising performance for images of vehicle plates as well as the enhancement of recognition accuracy. Interestingly, the pre-training-based approach did not achieve a significantly higher PSNR ($\sim 20.47$ dB) or SSIM ($\sim 0.64$) than the trained-from-scratch baseline (PSNR $\sim17.29$ dB, SSIM $\sim0.51$). However, this modest difference contributed to a drastic improvement for both character and sequence accuracy when applying LPRNet to the denoised images, showing the advantage of our pre-training framework. Therefore, the advances in training methods contributed to a diffractive denoiser that could become a viable solution to real-world applications such as traffic monitoring and surveillance in hard weather and low-light conditions.

The comparative analysis on the CBLPRD dataset demonstrated the superior data efficiency and convergence robustness afforded by our transfer learning framework. As shown in the Supplementary Figure S5, a baseline model trained from scratch on a limited subset (10K images) failed to resolve the structural details of the characters, resulting in blurred and unintelligible outputs (Supplementary Figure S5b). While increasing the training volume to 330K images improved the baseline's denoising performance, the resulting images still exhibited residual artifacts and lower contrast (Supplementary Figure S5c). Remarkably, the fine-tuned model (Supplementary Figure S5d), initialized with pre-trained weights and trained on only 10K images for 50 epochs, achieved reconstruction fidelity that surpassed even the large-scale baseline (330K). This performance leap was further elucidated by the visualization of the diffractive layer weights. The phase masks of the 10K baseline remained relatively smooth and lacked the necessary diffractive features to suppress complex noise. In contrast, the fine-tuned weights exhibited highly structured, high-frequency spatial modulations. These optimized phase distributions indicated that the model effectively retained the foundational denoising priors from the pre-training phase while rapidly adapting to the specific geometric morphology of alphanumeric and Chinese characters. This validated that our pre-training-enabled approach not only reduced the demand for large-scale domain-specific data but also reached a superior performance ceiling compared to conventional training-from-scratch methods.

\begin{figure}[thpb]
    \centering
    \includegraphics[width=\linewidth]{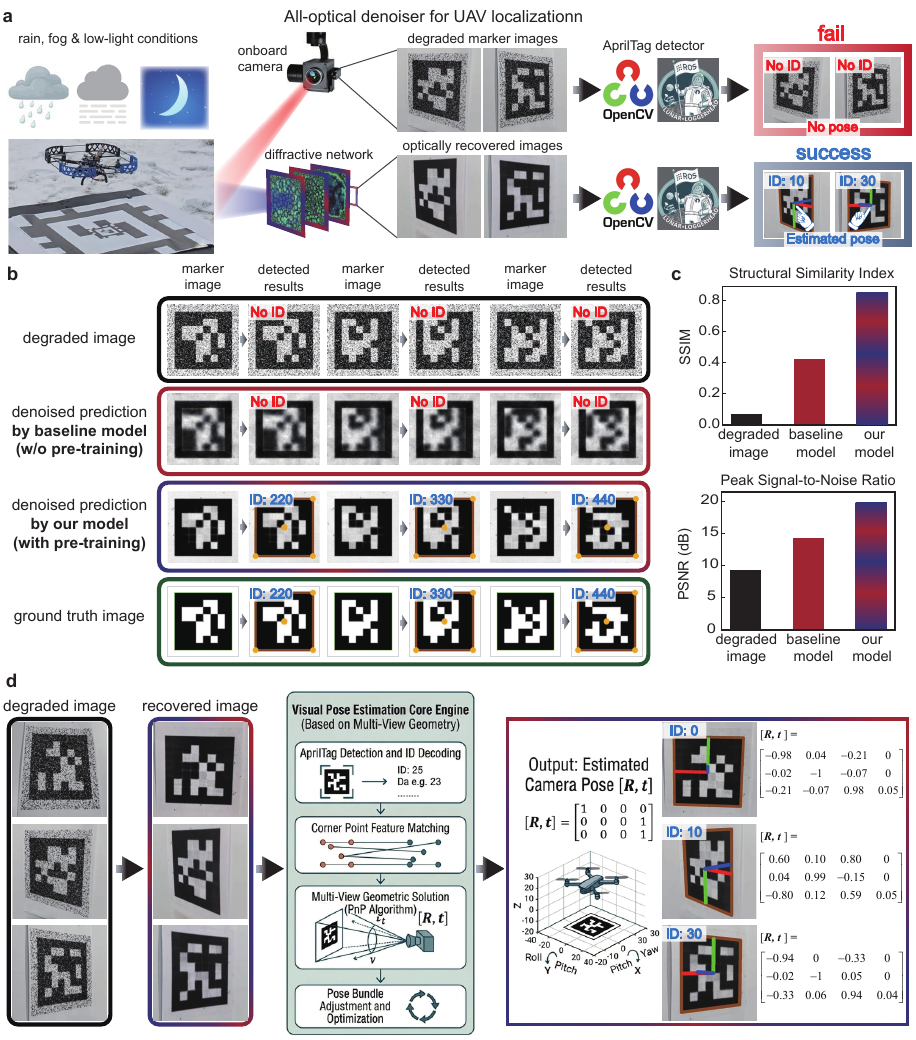}
    \caption{\textbf{Pre-training-enabled snapshot all-optical denoisers for UAV localization.} (a) Schematic of the pre-sensor diffractive neural network, which leverages pre-training and transfer learning to optimize optical preprocessing for robust UAV localization. (b) Comparative denoising performance illustrating the superior restoration of corrupted images by our diffractive processor compared to a baseline model trained from scratch. (c) Quantitative assessment of denoising and localization reliability on the AprilTag dataset, showing SSIM and PSNR metrics aggregated over 100 test images with diverse marker configurations. (d) The visual pose estimation pipeline, where high-fidelity AprilTag images restored by our diffractive model enable precise feature matching and PnP algorithm solutions to accurately determine the UAV's 6-DOF poses.}
    \label{F6}
\end{figure}

Figure \ref{F6}a further illustrated the application of our optical denoising model for UAV localization under challenging environmental conditions, such as low light and adverse weather. For this task, we demonstrated the capabilities of our pre-training-enabled optical model in denoising AprilTag markers, which served as critical visual anchors for autonomous navigation. Figure \ref{F6}b provided a visual comparison of the performance between our pre-training-enabled model and the baseline model (trained from scratch). The raw captured images under diverse noise conditions were severely degraded, leading to a complete failure of the standard detection algorithm (“No ID”). While the baseline model reduced some noise, it suffered from severe distortions and edge blur, still failing ID decoding. In contrast, our model achieved high-fidelity restoration of the tag's distinctive black-and-white patterns and sharp boundaries. This superior reconstruction directly enabled reliable tag detection, successful ID decoding, and accurate landmark localization. Figure \ref{F6}c provides a quantitative analysis of the denoising performance and the corresponding boost in localization robustness. Although the pre-training-enabled approach yielded relatively modest gains in pixel-wise metrics, with an average PSNR ($\sim 19.84$ dB) and SSIM ($\sim 0.86$) slightly higher than the baseline (PSNR $\sim 14.37$ dB, SSIM $\sim 0.42$), they translated into a decisive leap in pose estimation success rates. Finally, Figure \ref{F6}d detailed the visual pose estimation core engine, in which denoised outputs enable precise corner feature matching and PnP-based solutions, ultimately yielding accurate 6-DOF camera poses for stable UAV flight control in real-world environments. The final outputs confirmed the successful recovery of the 6-DOF trajectory and stable vehicle orientations, demonstrating that pre-training-enabled diffractive denoiser was a viable solution for enabling safe and accurate UAV operations in all-weather, all-time conditions.

The comparative results on the AprilTag dataset (Supplementary Figure S6) further underscored the transformative impact of our transfer learning approach, particularly in scenarios with extremely limited data. For this small-scale dataset (586 images), the baseline model (Supplementary Figure S6d) trained from scratch failed to reconstruct the essential structural integrity of the markers, producing blurry outputs where the unique ID coding and square boundaries are virtually unresolvable. Such degradation would lead to a total failure in downstream UAV localization tasks. In stark contrast, the fine-tuned model (Supplementary Figure S6c), benefiting from the robust denoising priors of the pre-trained network, achieved near-perfect restoration of the sharp geometric edges and high-contrast patterns required for reliable detection. The visualization of the diffractive layers revealed that the baseline’s phase masks were underdeveloped, lacking the intricate diffractive features necessary for precise wavefront modulation. However, the fine-tuned phase masks exhibited sophisticated, high-frequency spatial distributions that were effectively inherited and refined from the pre-trained phase. This allowed the model to achieve a superior performance ceiling with minimal task-specific data, validating that knowledge transfer was not only beneficial but essential for deploying diffractive processors in specialized, data-scarce applications like autonomous navigation and robotics.

Besides the above tasks, we also proceeded with the application of our optical denoising model for downstream image classification. Supplementary Figure S13a and S13b demonstrated that our model effectively concentrated optical energy into the correct detector regions, which are otherwise scattered by environmental noise. This enhancement was visually confirmed in Supplementary Figure S13c and S13d, where the restored features from diverse datasets enable both DNN and ANN classifiers to produce high-contrast focal spots for accurate class identification, resulting in a significant, dataset-wide increase in classification accuracy.

\section{Conclusions}\label{sec3}
Overall, we developed a powerful pre-training and knowledge-transfer approach that revolutionized the optimization of free-space optical networks, enabling single-shot all-optical image denoising with unprecedented fidelity. In particular, we demonstrated that the pre-training using a massive dataset of 3.45 million simple but diverse images prepared the diffractive network with a significantly better prior for image denoising tasks. Compared to the conventional train-from-scratch approach, the task-specific diffractive processors fine-tuned from this pre-trained model showed significantly superior performances (with up to 0.44 and 5.47 dB improvements for SSIM and PSNR, respectively) for denoising broad styles of images, including CIFAR-10, MedMNIST, CelebA, CBLPRD, and AprilTag. Specifically, the denoised images by our diffractive processor recovered well-defined features with suppressed residual background noise from noisy images up to $\gamma=0.5$. The pre-training process also allowed the effective generalization towards different types of corruptions, including Poisson and salt \& pepper noises as well as their mixtures, making the approach applicable to diverse applications. Physical implementation in a reflective cavity-type diffractive processor, realized via a phase-only spatial light modulator (SLM) and mirror-assisted folding for multi-layer diffraction, confirms the experimental viability of our pre-training-enabled architecture. The empirical results demonstrated that the processor effectively restored the primary morphological features of previously unseen noisy EMNIST characters, exhibiting high-fidelity agreement with numerical simulations. This robust correspondence between theory and experiment validated the system’s capacity for reliable wavefront reconstruction, positioning the diffractive neural network as a compact and high-speed optical front-end for real-time machine vision and intelligent sensing applications. The superiority of our diffractive denoiser enabled, for the first time, snapshot optical models as the preprocessing module for technologies ranging from face detection, vehicle plate recognition, and UAV localization. Moving forward, we believe that the pre-training framework holds tremendous value for developing next-generation optical and optoelectronic platforms with competitive inference and generalization performances. The all-optical diffractive denoiser we developed can also find broad applications in vision systems as a fast, pre-sensor information processing tool. 

\section{Methods}\label{sec4}
\subsection{Forward model of the all-optical diffractive image denoiser}\label{sec4_subsec1}
The all-optical diffractive denoiser consists of a sequence of diffractive layers, optimized through deep learning to perform specific wavefront modulations. In this physical framework, the light transport is governed by free-space propagation between: (i) the input plane to the initial diffractive surface, (ii) successive diffractive surfaces, and (iii) the final diffractive layer to the output detector plane. The propagation of the complex optical field in air is modeled using the angular spectrum method, derived from the Rayleigh–Sommerfeld diffraction integral \cite{lin2018all}:
\begin{equation}
    \begin{aligned}
{{U}_{out}}(x,y,z+d)={{\mathcal{F}}^{-1}}\{\mathcal{F}\{{{U}_{in}}(x,y,z)\}\cdot H({{f}_{x}},{{f}_{y}};d)\},
    \end{aligned}
\end{equation}
where ${{U}_{in}}$ denotes the input optical field, ${{U}_{out}}(x,y,z)$ is the optical field after propagation over an axial distance of $d$, operator $\mathcal{F}$ and ${{\mathcal{F}}^{-1}}$ represent the 2D Fourier transform and inverse Fourier transform respectively, $H({{f}_{x}},{{f}_{y}};d)$ is the free-space transfer function and defined as:
\begin{equation}
    \begin{aligned}
    H({{f}_{x}},{{f}_{y}};d)=\left\{ \begin{matrix}
    \begin{aligned}
       &\exp \left ( jkd\sqrt{1-{{\left( \frac{2\pi {{f}_{x}}}{k} \right)}^{2}}-{{\left( \frac{2\pi {{f}_{y}}}{k} \right)}^{2}}} \right),\qquad f_{x}^{2}+f_{y}^{2}\le \frac{1}{{{\lambda }^{2}}}  \\
       &0,\qquad \qquad \qquad \qquad \qquad \qquad \qquad \qquad \qquad \quad f_{x}^{2}+f_{y}^{2}>\frac{1}{{{\lambda }^{2}}}  \\
    \end{aligned}
    \end{matrix} \right.,
    \end{aligned}
\end{equation}
where ${{f}_{x}}$ and ${{f}_{y}}$ represents the spatial frequencies along the x-direction and y-direction, $\lambda $ is the illumination wavelength, $j=\sqrt{-1}$, and $k=\frac{2\pi }{\lambda }$ is the wavevector. Each diffractive layer is configured to provide pure phase modulation. The complex transmittance coefficient of the $l-th$ diffractive layer at coordinates $(x,y)$ is expressed as:
\begin{equation}
    \begin{aligned}
    {{t}^{l}}(x,y)=\exp \left( j{{\phi }^{l}}(x,y) \right),
    \end{aligned}
\end{equation}
where ${{\phi }^{l}}(x,y)$ represents the modulation of the optical field phase by the neuron at $(x,y)$ on the diffractive layer. By iteratively applying the propagation and modulation functions, the final complex field distribution after an arbitrary number of diffractive layers is analytically determined.

\subsection{Training loss functions}\label{sec4_subsec2}
The phase modulation coefficients of each diffractive layer are determined through a deep-learning-based optimization process. This involves iteratively refining the network parameters by (i) evaluating a loss function that quantified the discrepancy between the output intensity and the ground-truth pattern, and (ii) updating the phase distributions via backpropagation. To optimize the diffractive network for image denoising, we employ the normalized mean squared error (NMSE) as the objective function, defined as:
\begin{equation}
    \begin{aligned}
    {{\mathcal{L}}_{NMSE}}=\frac{1}{N}{{\sum\limits_{x,y}{\left( \frac{O(x,y)}{\max (O(x,y))}-\frac{G(x,y)}{\max (G(x,y))} \right)}}^{2}},
    \end{aligned}
\end{equation}
where $N$ denotes the total number of pixels in each image. The normalization of both the output optical field $O(x,y)$ and the ground truth intensity pattern $G(x,y)$, ensures they are mapped to an identical dynamic range. This approach prevents the loss calculation from being dominated by global intensity fluctuations or variations in total optical power, thereby focusing the optimization on the reconstruction of high-fidelity spatial features.

In the all-optical diffractive image denoiser, network performance obtained using only NMSE as the loss function is undesirable. To preserve structural correlations and fine textures, and enforce smoothness piecewise after denoising, we further induces two additional constraints, the Pearson correlation coefficient (PCC) loss and Total Variation (TV) loss. The PCC loss is defined as:
\begin{equation}
    \begin{aligned}
    {{\mathcal{L}}_{PCC}}=1-\frac{\sum\limits_{x,y}{\left( O(x,y)-\overline{O} \right)\left( G(x,y)-\overline{G} \right)}}{\sqrt{\sum\limits_{x,y}{{{\left( O(x,y)-\overline{O} \right)}^{2}}{{\left( G(x,y)-\overline{G} \right)}^{2}}}}},
    \end{aligned}
\end{equation}
where $\overline{O}$ and $\overline{G}$ represents the mean values of the output optical field and ground truth optical field, respectively. The TV loss is defined as:
\begin{equation}
    \begin{aligned}
    {{\mathcal{L}}_{TV}}=\sum\limits_{x,y}{\left( \left| O(x+1,y)-O(x,y) \right|+\left| O(x,y+1)-O(x,y) \right| \right)},
    \end{aligned}
\end{equation}

With the combination of the PCC loss and TV loss, the image denoiser could achieve a performance balance: the denoised image is clean yet retained its natural texture and structural coherence. The final loss function is defined as:
\begin{equation}
    \begin{aligned}
    \mathcal{L}={{w}_{NMSE}}{{\mathcal{L}}_{NMSE}}+{{w}_{PCC}}{{\mathcal{L}}_{PCC}}+{{w}_{TV}}{{\mathcal{L}}_{TV}},
    \end{aligned}
\end{equation}
where ${{w}_{NMSE}}$, ${{w}_{PCC}}$, and ${{w}_{TV}}$ represent different loss weighting factors assigned to NMSE, PCC, and TV loss.

\subsection{Preparation of image datasets}\label{sec4_subsec3}
In our numerical results, the diffractive neural network is initially pre-trained on the large-scale QuickDraw dataset, encompassing 3.45 million images across all 345 categories (10,000 images per category). This extensive repository is partitioned into a training set of 3 million images, a validation set of 100,000 images, and a test set of 350,000 images. To ensure robust feature learning, input images underwent a standardized preprocessing pipeline: raw samples were normalized, followed by bilinear interpolation to $150\times150$ pixels and zero-padding to a final dimension of $400\times400$ pixels. During optimization, we implements diverse data augmentation strategies—including random flipping, translation, scaling, cropping, and adjustments to brightness and contrast—to enhance the model's representative capacity. The random gaussian noise (noise probability $\gamma=0.5$) is added to the raw input images prior to the normalization process. To rigorously evaluate the generalization and zero-shot denoising capabilities of the pre-trained model, we introduce a diverse battery of noise profiles into the test datasets, including Gaussian, Salt-and-Pepper, and Poisson noise, as well as a more challenging hybrid configuration, Mixed noise (a composite of Gaussian, Salt-and-Pepper, and Poisson noise). The zero-shot generalization of this pre-trained model is subsequently evaluated using the MNIST (10,000 images), EMNIST (20,800 images), and Fashion-MNIST (10,000 images) benchmarks.

To investigate the efficacy of transfer learning, the pre-trained diffractive processor is fine-tuned and validated across five specialized datasets representing distinct complex domains. These includes CIFAR-10 (55,000 training/5,000 test images), widely utilized for conventional computer vision benchmarks, and medical imaging datasets from the MedMNIST collection, specifically ChestMNIST (80,000 training/9,687 test) and BloodMNIST (25,000 training/4,081 test). Furthermore, the model's performance is extended to high-stakes downstream tasks using the CelebA face dataset (90,000 training/1,439 test), the CBLPRD vehicle license plate dataset (330,000 training/1,110 test), and the AprilTag marker dataset for UAV localization (480 training/107 test). To assess the network's scalability for high-complexity scenes, we further constructs composite datasets by tiling ChestMNIST images into grids, with training/test splits of 8,000/1,000 and 3,200/300 images, respectively.

For the experimentally demonstrated design, the same partitioning of the QuickDraw dataset is maintained. To adapt to the physical apertures of the diffractive layers, the input images underwent a distinct multi-stage spatial transformation pipeline. Raw samples are first interpolated to $80\times80$ pixels and subjected to stochastic noise addition and normalization. Subsequently, to match the physical sampling requirements and the active area of the diffractive processor, the preprocessed images are further interpolated to 320×320 pixels and zero-padded to a final input dimension of $900\times900$ pixels. This hierarchical scaling strategy ensures that the underlying image features are effectively mapped onto the diffractive neurons while maintaining sufficient spatial resolution for high-fidelity optical denoising in real-world scenarios.

\subsection{Evaluation metrics}\label{sec4_subsec4}
To quantitatively evaluate the denoising performance, we employ the peak signal-to-noise ratio (PSNR) and the structural similarity index (SSIM) \cite{wang2004image}. The PSNR measures the reconstruction fidelity between the ground-truth image $G(x,y)$ and the denoised counterpart output $O(x,y)$, defined as:
\begin{equation}
    \begin{aligned}
    PSNR=10{{\log }_{10}}(\frac{1}{\frac{1}{N}\sum\nolimits_{x,y}{{{\left| O(x,y)-\frac{\sum\nolimits_{x,y}{O(x,y)\cdot G(x,y)}}{\sum\nolimits_{x,y}{G{{(x,y)}^{2}}}}\cdot G(x,y) \right|}^{2}}}}),
    \end{aligned}
\end{equation}
where $N$ denotes the total number of pixels. Higher PSNR values correspond to a more accurate restoration of signal integrity relative to noise. Complementary to PSNR, the SSIM provides an assessment of the perceptual quality by measuring the preservation of structural information. Unlike pixel-wise metrics, SSIM accounts for luminance, contrast, and structure across local image patches, expressed as:
\begin{equation}
    \begin{aligned}
    SSIM(G,O)=\frac{(2{{\mu }_{G}}{{\mu }_{O}}+{{C}_{1}})(2{{\sigma }_{G,O}}+{{C}_{2}})}{(\mu _{G}^{2}+\mu _{O}^{2}+{{C}_{1}})(\sigma _{G}^{2}+\sigma _{O}^{2}+{{C}_{2}})},
    \end{aligned}
\end{equation}
where ${{\mu }_{G}}$ and ${{\mu }_{O}}$ denotes the mean value of image $G$ and $O$, ${{\mu }_{G}}$ and ${{\mu }_{O}}$ represents the corresponding standard deviation, ${{\sigma }_{G,O}}$ is the covariance between the two patterns, and ${{C}_{1}}, {{C}_{2}}$ are stabilization constants. This metric quantifies the consistency of structural features, with a value of unity representing an identical reconstruction.

\subsection{Implementation details for the numerical results}\label{sec4_subsec5}
In the numerical simulation, the minimum feature size of each diffractive layer is constrained to $0.47\lambda$, while the physical pixel pitch for both input and output planes is set at $0.94\lambda$. The diffractive image denoiser consists 6 layers, each containing $270\times270$ independent modulation units. To ensure high-fidelity modeling of the optical field, the computational grid resolution is refined to $0.235\lambda$, allowing each minimum feature to be resolved by a $2\times2$ grid point matrix. Furthermore, the computational domain is expanded via zero-padding to $400\times400$ points to mitigate boundary artifacts and enhance the precision of the angular spectrum propagation.

The network parameters are optimized using the Adam optimizer with an initial learning rate of 0.01 and a batch size of 9,600. All simulations are implemented in Python (v3.10.19) using the PyTorch (v2.2.0) library, executed on a high-performance workstation equipped with 8 NVIDIA A100 GPUs (40 GB) and dual Intel 8358P CPUs. To balance gradient exploration and convergence stability, we employ a Warmup-Stable-Decay (WSD) \cite{hagele2024scaling} learning rate scheduling scheme. During the initial ${{E}_{w}}$ warmup epochs, the learning rate increases linearly from a predefined minimum value ${{\eta }_{\operatorname{m}}}$ to the target base ${{\eta }_{b}}$ to suppress early-stage gradient volatility and prevent premature convergence to suboptimal local minima. This is followed by a stable phase of ${{E}_{s}}$ epochs at a constant high learning rate to accelerate convergence. In the final ${{E}_{d}}$ epochs, the learning rate is systematically decayed, facilitating precise fine-tuning of the phase coefficients near the global loss minimum. The above WSD learning rate adjustment scheduler is illustrated as follows:
\begin{equation}
    \begin{aligned}
    {{\eta }_{e}}=\left\{ \begin{aligned}
      & {{\eta }_{\operatorname{m}}}+\frac{e}{{{E}_{w}}}\cdot ({{\eta }_{b}}-{{\eta }_{\operatorname{m}}}),\qquad\qquad\qquad\qquad\qquad\qquad\qquad e<{{E}_{w}} \\ 
     & {{\eta }_{b}},\qquad\qquad\qquad\qquad\qquad\qquad\qquad\qquad\qquad\qquad\qquad {{E}_{w}}\le e<{{E}_{w}}+{{E}_{s}} \\ 
     & {{\eta }_{\operatorname{m}}}+\frac{1}{2}({{\eta }_{b}}-{{\eta }_{\operatorname{m}}})\left( 1+\cos \left( \frac{\pi (e-{{E}_{w}}-{{E}_{s}})}{{{E}_{t}}-{{E}_{w}}-{{E}_{s}}} \right) \right),\qquad e\ge {{E}_{w}}+{{E}_{s}} \\ 
    \end{aligned} \right.,
    \end{aligned}
\end{equation}
where ${{\eta }_{e}}$ represents the learning rate at epoch $e$, ${{E}_{t}}$ denotes the total number of the training epochs. For the numerically-tested designs, ${{E}_{t}}$, ${{E}_{s}}$, and ${{E}_{w}}$ are 500, 100, and 50 epochs, respectively.

To extend the capabilities of the pre-trained diffractive model to specialized complex datasets, we implement a robust fine-tuning framework that integrates second-order stochastic optimization (Sophia) \cite{liusophia} with weight interpolation (WiSE-FT) \cite{wortsman2022robust}. This two-stage strategy is designed to balance task-specific adaptation with out-of-distribution robustness. During the fine-tuning phase, the Sophia optimizer is employed to leverage Hessian diagonal estimates for dynamic, parameter-wise learning rate adjustment. The parameter update at step $t$ is governed by:
\begin{equation}
    \begin{aligned}
    {{\theta }_{t+1}}={{\theta }_{t}}-\eta \cdot clip(\frac{{{g}_{t}}}{\max ({{h}_{t}},\varepsilon )},\rho ),
    \end{aligned}
\end{equation}
where ${{g}_{t}}$ denotes the gradient, ${{h}_{t}}$ represents the exponential moving average of the diagonal Hessian estimate, $\rho $ is a clipping threshold utilized to stabilize updates against sharp curvature. Following the optimization of the fine-tuned weights, ${{\theta }_{FT}}$, we apply a linear interpolation with the initial pre-trained weights, ${{\theta }_{PT}}$, to derive the final model parameters:
\begin{equation}
    \begin{aligned}
    \theta =\alpha \cdot {{\theta }_{PT}}+(1-\alpha )\cdot {{\theta }_{FT}},
    \end{aligned}
\end{equation}
where the mixing coefficient $\alpha \in [0,1]$ modulates the trade-off between the broad generalization of the pre-trained model and the specificity required for the target task. This approach preserves the rapid convergence and curvature awareness inherent in second-order optimization while utilizing the WiSE-FT protocol to mitigate catastrophic forgetting and enhance resilience to domain shifts. For all transfer learning tasks, the fine-tuning duration is standardized to 50 epochs.

\subsection{Implementation details for the experimental results}\label{sec4_subsec6}
To better fit the experimental model, we introduce an additional diffraction-efficiency loss term ${{\mathcal{L}}_{diff}}$:
\begin{equation}
    \begin{aligned}
        \begin{aligned}
          & {{\mathcal{L}}_{diff}}=\left\{ \begin{aligned}
          & {{e}^{-r}},\quad r\le {{r}_{eff}} \\ 
         & 0,\qquad r>{{r}_{eff}} \\ 
        \end{aligned} \right. \\ 
         & r=\frac{{{P}_{out}}}{{{P}_{in}}} \\ 
        \end{aligned}
    \end{aligned}
\end{equation}
where ${{P}_{in}}$ and ${{P}_{out}}$ represents the total optical intensity on the input FoV and the output FoV, respectively, ${{r}_{eff}}$ is the threshold efficiency ratio. In this experiment, ${{r}_{eff}}$ is empirically set to 30\% and the loss function ${{\mathcal{L}}_{diff}}$ is activated only when the diffraction efficiency fell below this threshold.

To account for the oblique incidence of the beam on the Spatial Light Modulator (SLM), we replace conventional angular spectrum propagation with an off-axis angular spectrum method. This approach ensures an accurate physical modeling of tilted-beam diffraction in free space, which can be expressed as:
\begin{equation}
    \begin{aligned}
      & {{G}_{off}}({{f}_{xs}},{{f}_{ys}};z)=G({{f}_{xs}}+{{f}_{x0}},{{f}_{ys}}+{{f}_{y0}};0) \cdot \\ 
     & \exp \left\{ j2\pi ({{f}_{xs}}{{x}_{0}}+{{f}_{ys}}{{y}_{0}})+jkz\sqrt{1-\frac{4{{\pi }^{2}}}{{{k}^{2}}}\left[ {{({{f}_{xs}}+{{f}_{x0}})}^{2}}+{{({{f}_{ys}}+{{f}_{y0}})}^{2}} \right]} \right\} \\ 
    \end{aligned}
\end{equation}
where ${{G}_{off}}({{f}_{xs}},{{f}_{ys}};z)$ represents the off-axis spatial frequency spectrum, ${{f}_{xs}}$ and ${{f}_{ys}}$ correspond to the displaced spatial frequencies along the x and y directions, respectively, while ${{f}_{x0}}$ and ${{f}_{y0}}$ denote the spatial domain carrier frequencies induced by the tilt. This off-axis treatment is critical for maintaining phase fidelity in the non-paraxial regime typical of tilted optical architectures.

In our experimental implementation, the SLM pixel pitch of $3.6 \mu m$ serves as the fundamental sampling grid for the numerical simulations. The diffractive denoiser is architected with three layers, each comprising $350\times350$ neurons. To maintain physical consistency, each neuron is mapped to a $2\times2$ block of SLM pixels, yielding an effective lateral dimension of $7.2\mu m$, which is uniformly applied across the input and output planes. Geometric alignment is ensured by zero-padding the input/output images to match the layer dimensions, supplemented by a 100-pixel marginal buffer around each diffractive layer to mitigate boundary artifacts and align simulation parameters with the experimental clear aperture. Precision in the physical assembly is achieved through a multi-degree-of-freedom alignment system: two motorized rotation stages provide fine-grained angular control over the SLM and the reflecting mirror, while independent three-axis translation stages ensure axial and height synchronization. This rigorous optomechanical configuration guarantee that the optical beam accurately address the designated diffractive zones on the SLM surface during multiple reflections. Collectively, these measures establish a high-fidelity correspondence between the computational model and the experimental prototype, providing a robust foundation for validating the diffractive network's optical performance.

\bmhead{Acknowledgements}
This research was supported by the the National Natural Science Foundation of China (grant no. 62535008, 62405076, 12404442, 62335005, 12334016, 12025402, 62125501, 12261131500 and 92250302); National Key Research and Development Program of China (grant no. 2024YFB2809200, 2021YFA1400802 and 2022YFA1404700); the Guangdong Basic and Applied Basic Research Foundation (no. 2023A1515110685, 2025A1515011713); the Guangdong Provincial Key Laboratory of Semiconductor Optoelectronic Materials and Intelligent Photonic Systems(no. 2023B1212010003); the Guangdong Provincial Quantum Science Strategic Initiative (no. GDZX2306002); the Shenzhen Science and Technology Program (no. JCYJ20240813113603005, JCYJ20240813104929039, JCYJ20250604145557075); Shenzhen Fundamental research projects (no. JCYJ20241202123729038, JCYJ20241202123719025, JCYJ20220818102218040, GXWD20220817145518001).

\section*{Declarations}
The authors declare no competing financial interest.

\bibliography{ref}

\end{document}


\title[Article Title]{\textbf{\centerline{Supplementary Information for}}\\ 
Pre-training Enables Extraordinary All-optical Image Denoising}

\author[1,2]{\fnm{Xudong} \sur{Lv}}\email{lvxudong@hdu.edu.cn}
\equalcont{These authors contributed equally to this work.}

\author[1,3]{\fnm{Yuxiang} \sur{Sun}}\email{yuxiangsun@link.cuhk.edu.cn}
\equalcont{These authors contributed equally to this work.}

\author*[1]{\fnm{Shuo} \sur{Wang}}\email{wangshuo@hit.edu.cn}
\equalcont{These authors contributed equally to this work.}

\author[1]{\fnm{Nanxing} \sur{Chen}}\email{24b321003@stu.hit.edu.cn}

\author[3]{\fnm{Jun} \sur{Guan}}\email{guanjun@cuhk.edu.cn}

\author*[1,4,5]{\fnm{Jingtian} \sur{Hu}}\email{hujingtian@hit.edu.cn}

\affil[1]{\orgdiv{Ministry of Industry and Information Technology Key Lab of Micro-Nano Optoelectronic Information System, Guangdong Provincial Key Laboratory of Semiconductor Optoelectronic Materials and Intelligent Photonic System}, \orgname{Harbin Institute of Technology}, \orgaddress{\city{Shenzhen}, \postcode{518055}, \state{Guangdong}, \country{China}}}

\affil[2]{\orgdiv{School of Electronics and Information Engineering, Zhejiang Provincial Key Laboratory of Intelligent Vehicle Electronics Research}, \orgname{Hangzhou Dianzi University}, \orgaddress{\city{Hangzhou}, \postcode{310018}, \state{Zhejiang}, \country{China}}}

\affil[3]{\orgdiv{School of Science and Engineering}, \orgname{The Chinese University of Hong Kong (Shenzhen)}, \orgaddress{\city{Shenzhen}, \postcode{518172}, \state{Guangdong}, \country{China}}}

\affil[4]{\orgname{Quantum Science Center of Guangdong-Hong Kong-Macao Greater Bay Area}, \orgaddress{\city{Shenzhen}, \postcode{518055}, \state{Guangdong}, \country{China}}}

\affil[5]{\orgdiv{Key Laboratory of Photonic Technology for Integrated Sensing and Communication, Ministry of Education}, \orgname{Guangdong University of Technology}, \orgaddress{\city{Guangzhou}, \postcode{200124}, \state{Guangdong}, \country{China}}}

\maketitle

\textbf{This PDF file includes:}
Supplementary Figs. S1-S13

\section{Training process of the pre-trained model}\label{secA1}
\begin{figure}[thpb]
    \centering
    \includegraphics[width=\linewidth]{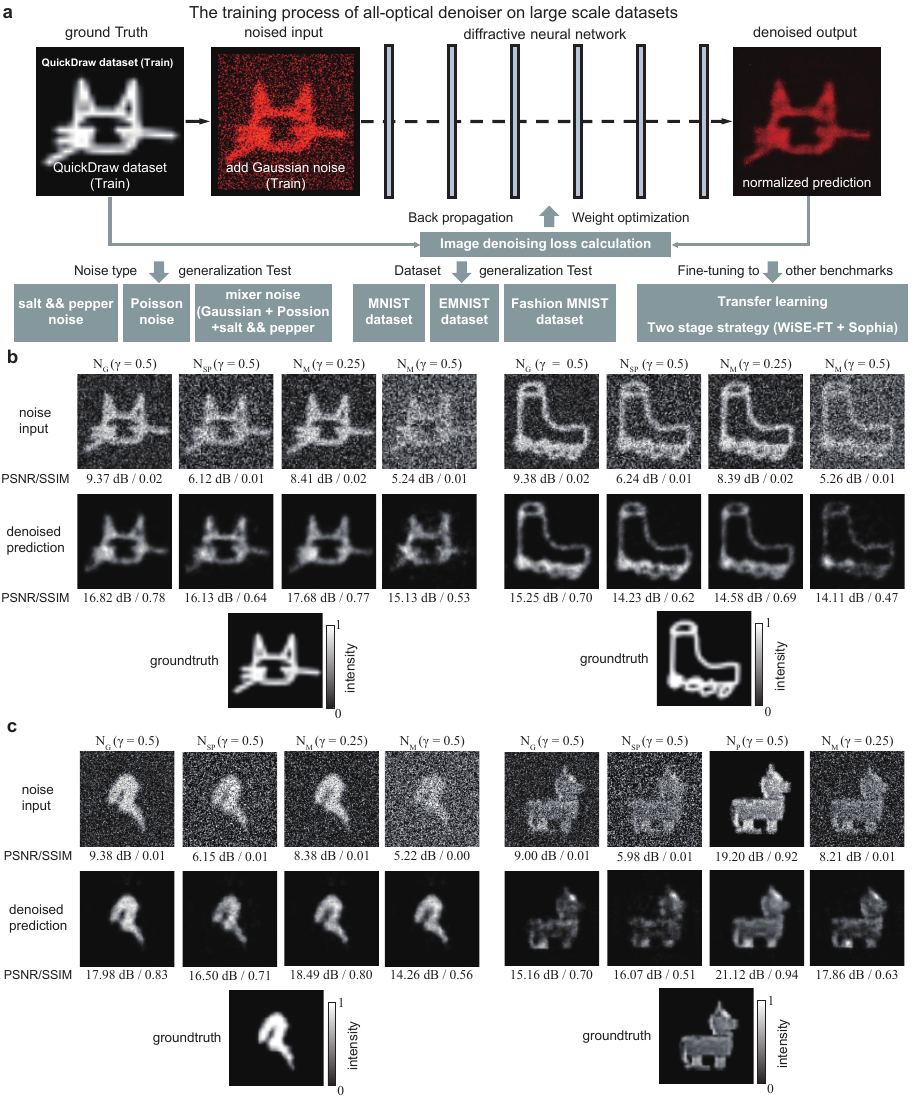}
    \caption{\textbf{The training process and zero-shot generalization testing of the pre-trained model.} (a) Schematic of the pre-training workflow, where the diffractive layers are iteratively optimized on a large-scale dataset (QuickDraw) to minimize the difference between the denoised output and the ground-truth. (b) Evaluation of the pre-trained model on previously unseen images from the same domain against various noise types, including Gaussian ${{N}_{G}}$, and Salt \& Pepper noise ${{N}_{SP}}$, as well as their mixture ${{N}_{M}}$ ($\gamma $ is the noise probability). (c) Validation of universal generalization across both novel datasets and unseen noise categories (including additional Poisson noise ${{N}_{P}}$).}
    \label{S1}
\end{figure}
The training process of the pre-trained model—along with the subsequent cross-domain generalization and fine-tuning on other benchmarks using a two-stage transfer learning strategy—was illustrated in Figure \ref{S1}. When tested on previously unseen images from the same domain, the pre-trained model demonstrated robust denoising performance against various types of interference, including Gaussian, Poisson, and salt-and-pepper noise, as well as hybrid combinations thereof. In parallel, we validated its universal generalization capability across both novel datasets and unseen noise categories. Notably, the diffractive processor maintained high-fidelity reconstruction when deployed on out-of-distribution datasets (EMNIST and Fashion-MNIST) corrupted by diverse noise sources. This performance hierarchy—spanning domain-specific optimization, cross-dataset adaptability, and multi-noise resilience—confirmed the effectiveness of our large-scale pre-training strategy in capturing universal structural priors for all-optical image restoration. The fine-tuning results on additional complex benchmarks are presented in the next section (Appendix \ref{secA2}).

\section{More results on other benchmarks}\label{secA2}
\begin{figure}[thpb]
    \centering
    \includegraphics[width=\linewidth]{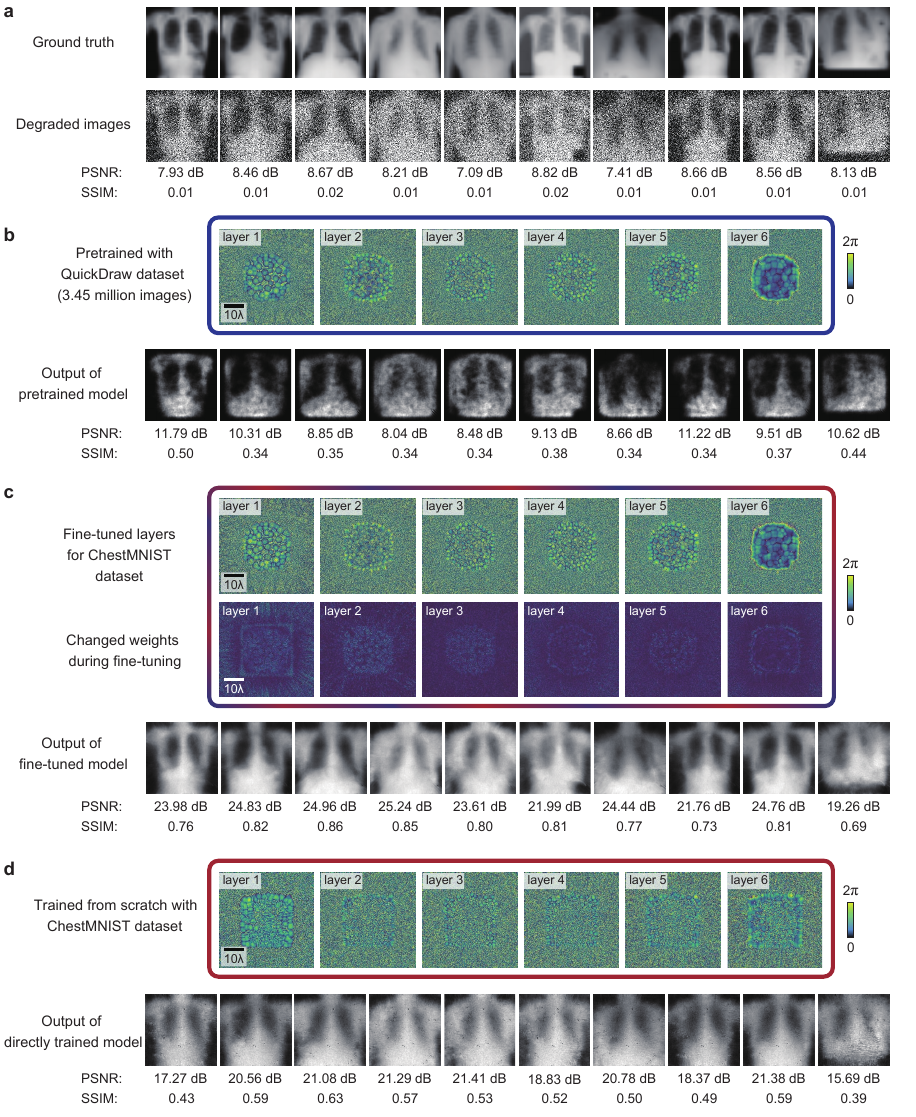}
    \caption{\textbf{Numerical tests on ChestMNIST dataset for diffractive denoiser obtained by knowledge transfer from a pre-trained model.} (a) Example images of the ground truth and input noise images from the test dataset based on ChestMNIST. Design and all-optical denoising output for diffractive networks optimized through (b) only pre-training (using the QuickDraw-based dataset), (c) pre-training and fine-tuning with ChestMNIST, and (d) direct training with ChestMNIST (i.e., the trained-from-scratch baseline model).}
    \label{S2}
\end{figure}

\begin{figure}[thpb]
    \centering
    \includegraphics[width=\linewidth]{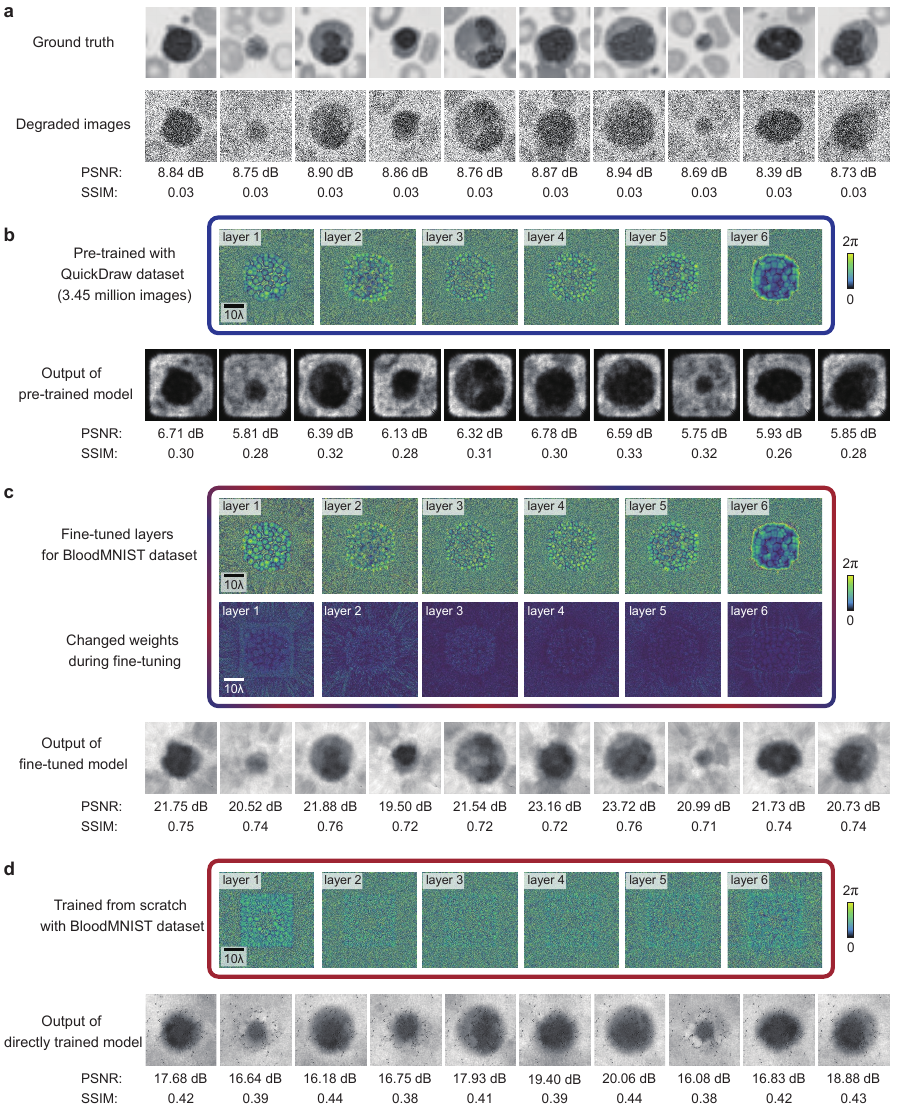}
    \caption{\textbf{Numerical tests on BloodMNIST dataset for diffractive denoiser obtained by knowledge transfer from a pre-trained model.} (a) Example images of the ground truth and input noise images from the test dataset based on BloodMNIST. Design and all-optical denoising output for diffractive networks optimized through (b) only pre-training, (c) pre-training and fine-tuning with BloodMNIST, and (d) direct training with BloodMNIST.}
    \label{S3}
\end{figure}

\begin{figure}[thpb]
    \centering
    \includegraphics[width=\linewidth]{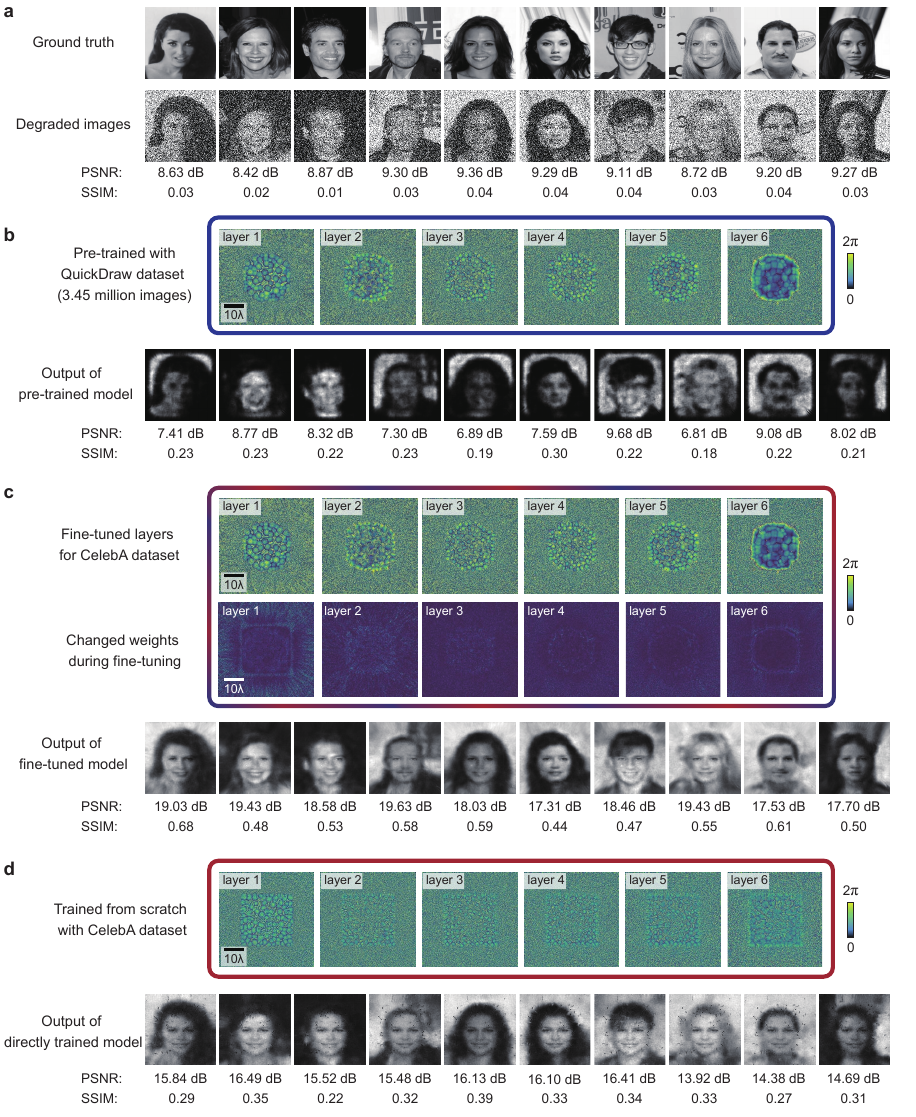}
    \caption{\textbf{Numerical tests on CelebA dataset for diffractive denoiser obtained by knowledge transfer from a pre-trained model.} (a) Example images of the ground truth and input noise images from the test dataset based on CelebA. Design and all-optical denoising output for diffractive networks optimized through (b) only pre-training, (c) pre-training and fine-tuning with CelebA, and (d) direct training with CelebA.}
    \label{S4}
\end{figure}

\begin{figure}[thpb]
    \centering
    \includegraphics[width=\linewidth]{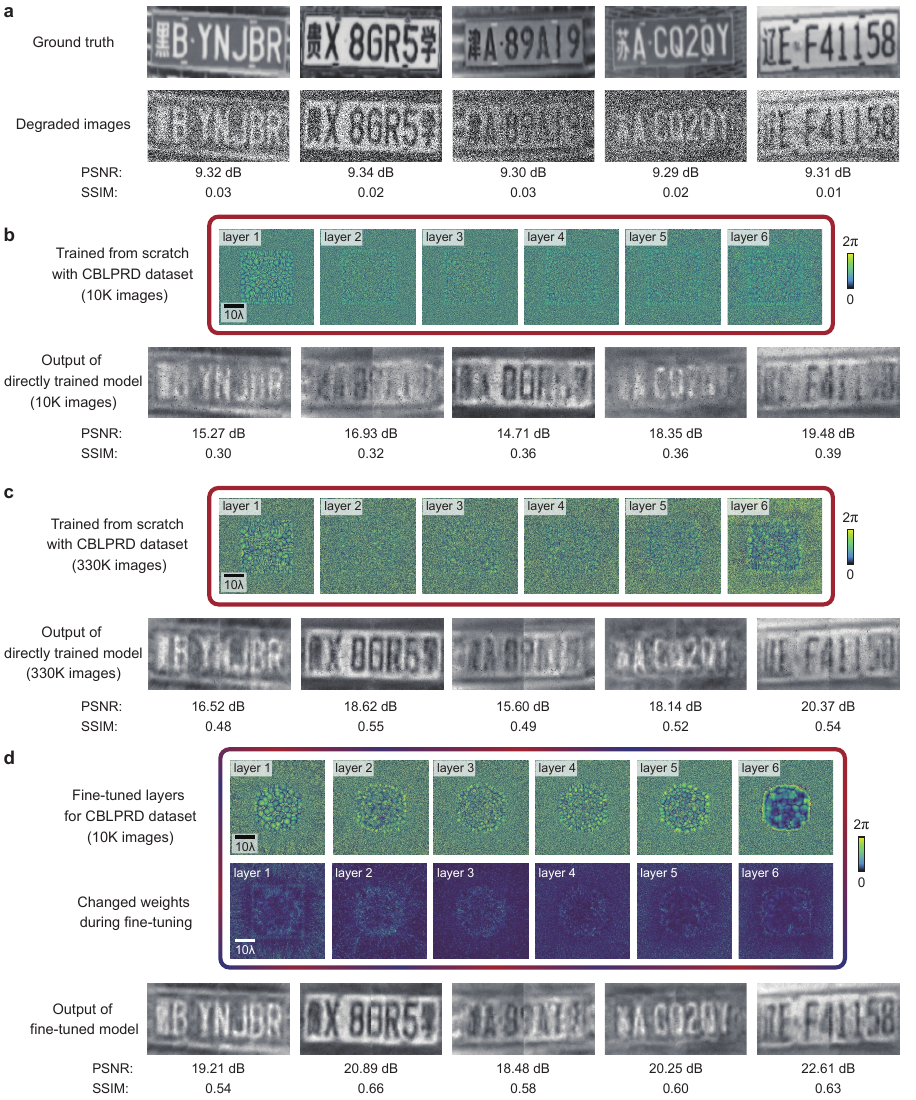}
    \caption{\textbf{Numerical tests on CBLPRD dataset for diffractive denoiser obtained by knowledge transfer (with different scales) from a pre-trained model.} (a) Example images of the ground truth and input noise images from the test dataset based on CBLPRD. Design and all-optical denoising output for diffractive networks optimized through (b) direct training with CBLPRD on small-scale sub-datasets (10K images), and (c) direct training with CBLPRD on large-scale datasets (330K images), and (d) pre-training and fine-tuning with CBLPRD.}
    \label{S5}
\end{figure}

\begin{figure}[thpb]
    \centering
    \includegraphics[width=\linewidth]{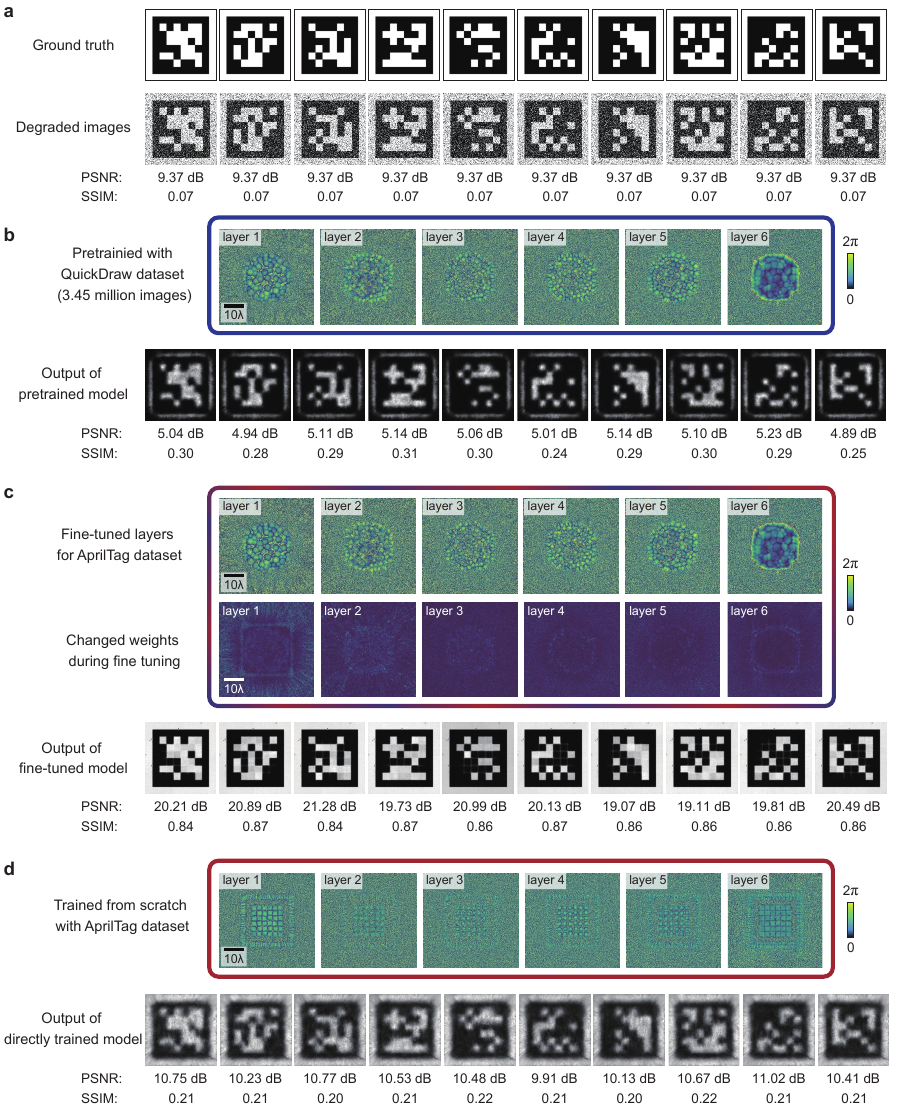}
    \caption{\textbf{Numerical tests on AprilTag dataset for diffractive denoiser obtained by knowledge transfer from a pre-trained model on data-constrained scenarios (586 images).} (a) Example images of the ground truth and input noise images from the test dataset based on AprilTag. Design and all-optical denoising output for diffractive networks optimized through (b) only pre-training, (c) pre-training and fine tuning with AprilTag, and (d) direct training with AprilTag.}
    \label{S6}
\end{figure}

The generalizability of our pre-trained diffractive visual processor was validated across a broad spectrum of complex datasets, including medical imaging (Figure \ref{S2} and \ref{S3}),  natural faces (Figure \ref{S4}), vehicle plates (Figure \ref{S5}), and robotic markers (Figure \ref{S6}). In each case, the fine-tuned model consistently outperformed the baseline models trained from scratch, particularly in recovering high-frequency details and maintaining structural integrity under severe noise interference. 

Specifically, for the vehicle plate (CBLPRD) dataset, we conducted a comparative scaling analysis to evaluate data efficiency. The results revealed that our transfer-learning-based model, fine-tuned with only 10K images, achieved superior denoising fidelity and semantic restoration compared to a baseline model trained from scratch on a massive dataset of 330K images. This significant reduction in data requirement—by a factor of 33—demonstrates that our pre-trained layers have successfully internalized universal structural priors, allowing for high-performance restoration in data-intensive tasks with minimal supervision. To further investigate the underlying mechanism of our transfer learning framework, we performed a comparative visualization of the optimized phase masks for the CBLPRD dataset. Intriguingly, the phase distributions of our model, fine-tuned on a small-scale subset of only 10K images, exhibited structural similarity and consistent color-mapped profiles to those of the baseline model trained from scratch on the full-scale dataset (330K images). This convergence in the spatial frequency domain indicated that the pre-training phase effectively established a robust set of universal diffraction primitives. By fine-tuning with a minimal data footprint, the diffractive layers were able to rapidly re-align their modulation weights to capture the domain-specific geometric features of vehicle plates, ultimately reaching a physical representational state nearly identical to the computationally expensive baseline.


\begin{figure}[thpb]
    \centering
    \includegraphics[width=0.99\linewidth]{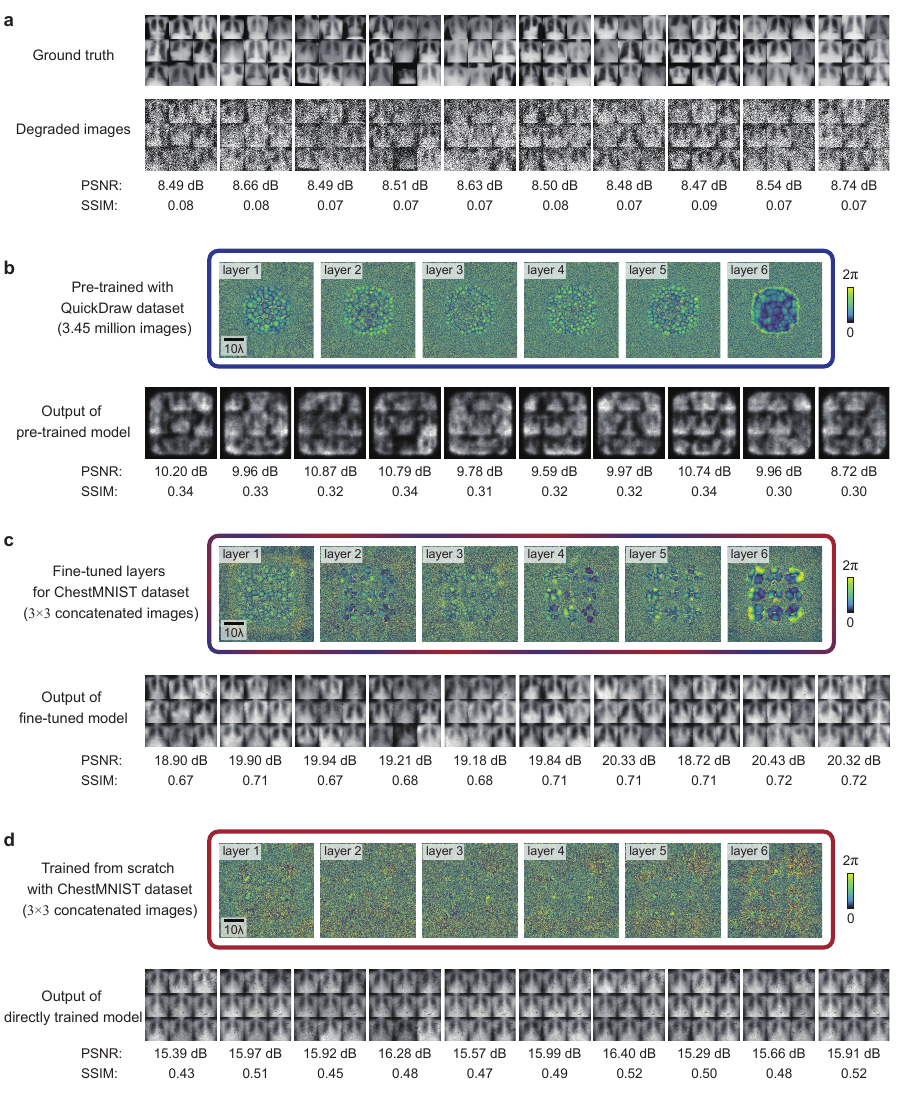}
    \caption{\textbf{Parallel processing of multiple images ($3\times3$) by scaled-up diffractive denoisers obtained by knowledge transfer from a pre-trained model.} (a) Example images of the ground truth and input noise images from the test dataset based on Concat-image ($3\times3$). Design and all-optical denoising output for diffractive networks optimized through (b) only pre-training, (c) pre-training and fine-tuning with Concat-image ($3\times3$), and (d) direct training with Concat-image ($3\times3$).}
    \label{S7}
\end{figure}

\begin{figure}[thpb]
    \centering
    \includegraphics[width=0.99\linewidth]{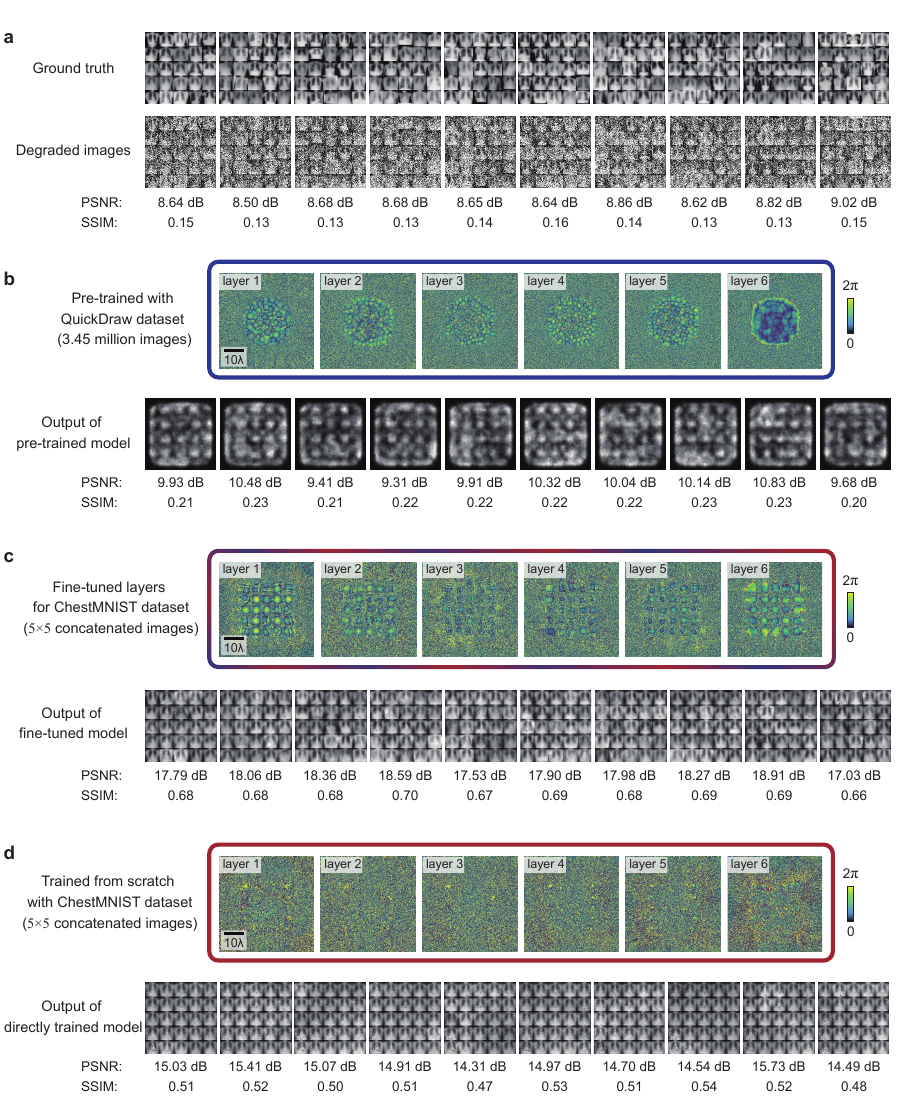}
    \caption{\textbf{Parallel processing of multiple images ($5\times5$) by scaled-up diffractive denoisers obtained by knowledge transfer from a pre-trained model.} (a) Example images of the ground truth and input noise images from the test dataset based on Concat-image ($5\times5$). Design and all-optical denoising output for diffractive networks optimized through (b) only pre-training, (c) pre-training and fine-tuning with Concat-image ($5\times5$), and (d) direct training with Concat-image ($5\times5$).}
    \label{S8}
\end{figure}

For AprilTag markers, which are critical for robotic localization and spatial intelligence, the processor demonstrated remarkable robustness even when fine-tuned on an extremely sparse dataset of only 586 images. Despite this minimal sample size, the model effectively reconstructed sharp geometric boundaries and high-contrast bit patterns essential for downstream decoding, whereas baseline methods failed to suppress significant artifacts. Furthermore, for medical and facial datasets, the processor successfully restored delicate anatomical textures and features that were otherwise obscured. The model also demonstrated remarkable scalability in processing composite scenes, as evidenced by the high-fidelity restoration of $3\times3$ (Figure \ref{S7}) and $5\times5$ (Figure \ref{S8}) tiled image mosaics. These results collectively underscored the efficacy of our transfer learning framework in capturing universal image priors, enabling the diffractive neural network to adapt to highly specialized and data-scarce domains with superior reconstruction precision.

\section{Performance comparison with 4-f filtering}\label{secA3}
\begin{figure}[thpb]
    \centering
    \includegraphics[width=0.95\linewidth]{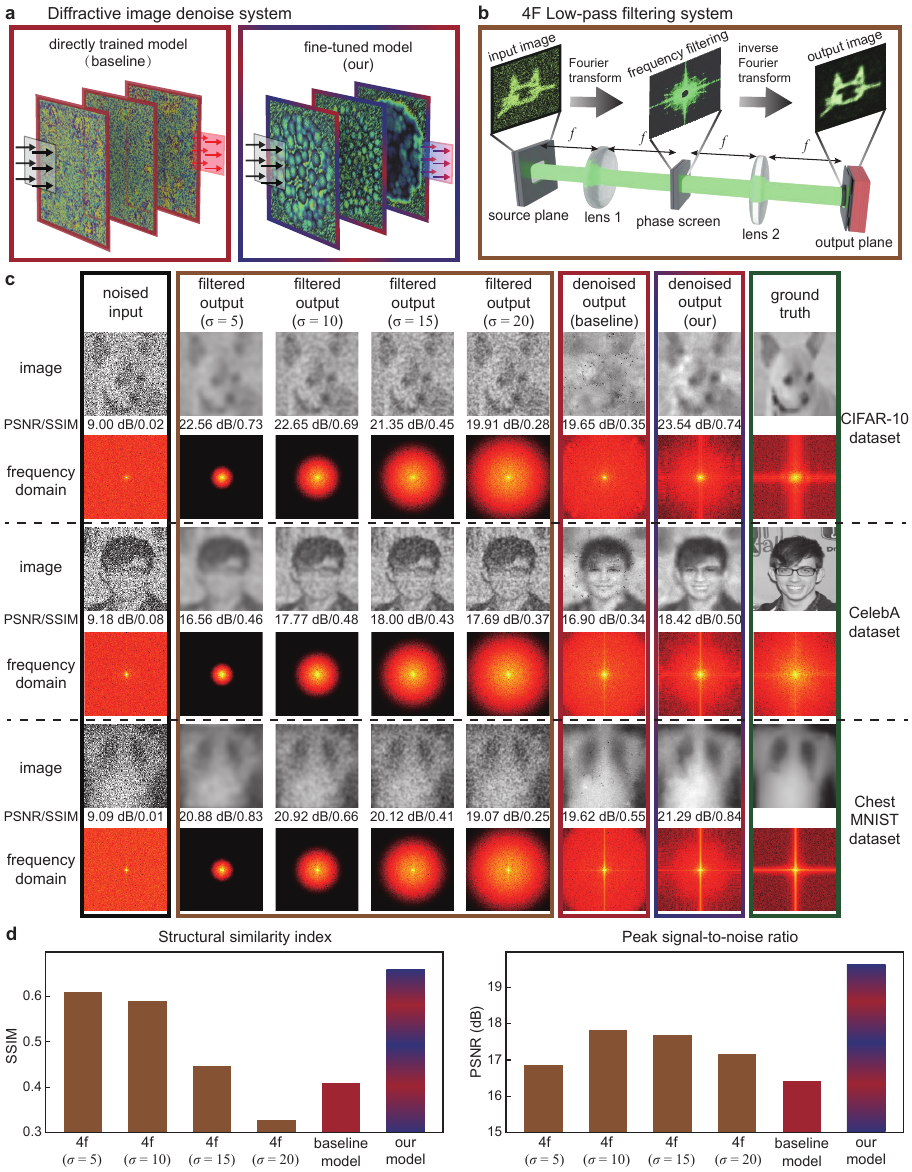}
    \caption{\textbf{Comparative analysis of 6-layer all-optical diffractive denoisers, a 4-f lens-based filtering system for filtering out Gaussian noise.} (a) Optical architectures of the 6-layer diffractive neural network, and (b) the lens-based 4-f filtering system. (c) Representative denoising results under a noise probability of 0.5. The performance of the optimized diffractive denoisers is benchmarked against 4-f systems configured with varying filtering ratios. (d) Quantitative evaluation of denoising efficacy across three distinct datasets. Performance metrics, including the SSIM and PSNR, were statistically aggregated over 1,000 independent test images per dataset.}
    \label{S9}
\end{figure}

The comparative results across diverse datasets, including natural scenes (CIFAR-10), facial portraits (CelebA), and medical imaging (ChestMNIST), highlighted the superior performance of our diffractive visual processor in contrast to conventional lens-based 4-f filtering systems (Figure \ref{S9}). While 4f systems utilizing spatial low-pass filters can effectively suppress high-frequency stochastic noise, they inherently suffered from a fundamental and irreversible trade-off between noise attenuation and structural preservation. As evidenced by the power spectra and reconstruction results, the removal of noise in a 4-f configuration leaded to a significant loss of sharp edges and fine anatomical details, characterized by truncated frequency support and localized blurring. In stark contrast, our optimized diffractive denoiser—leveraging learned multi-layer phase modulations—achieved high-fidelity restoration of both global contours and intricate textures. The preserved broad frequency characteristics in the output spectra demonstrated that the diffractive layers have internalized sophisticated, non-local wavefront processing logic that transcended the limitations of simple linear Fourier filtering. This was further corroborated by the corresponding spatial frequency distributions, where our model maintained a significantly wider spectral bandwidth while effectively isolating signal from interference. Quantitative metrics, specifically the PSNR and SSIM, provided rigorous validation of this advantage. Across all tested benchmarks, the diffractive approach consistently yielded higher reconstruction accuracy and structural fidelity than any fixed-aperture 4-f filter, even under extreme noise conditions, underscoring the efficacy of task-specific optical representation learning in complex image restoration.

\section{Analysis on noise probability}\label{secA4}
\begin{figure}[thpb]
    \centering
    \includegraphics[width=\linewidth]{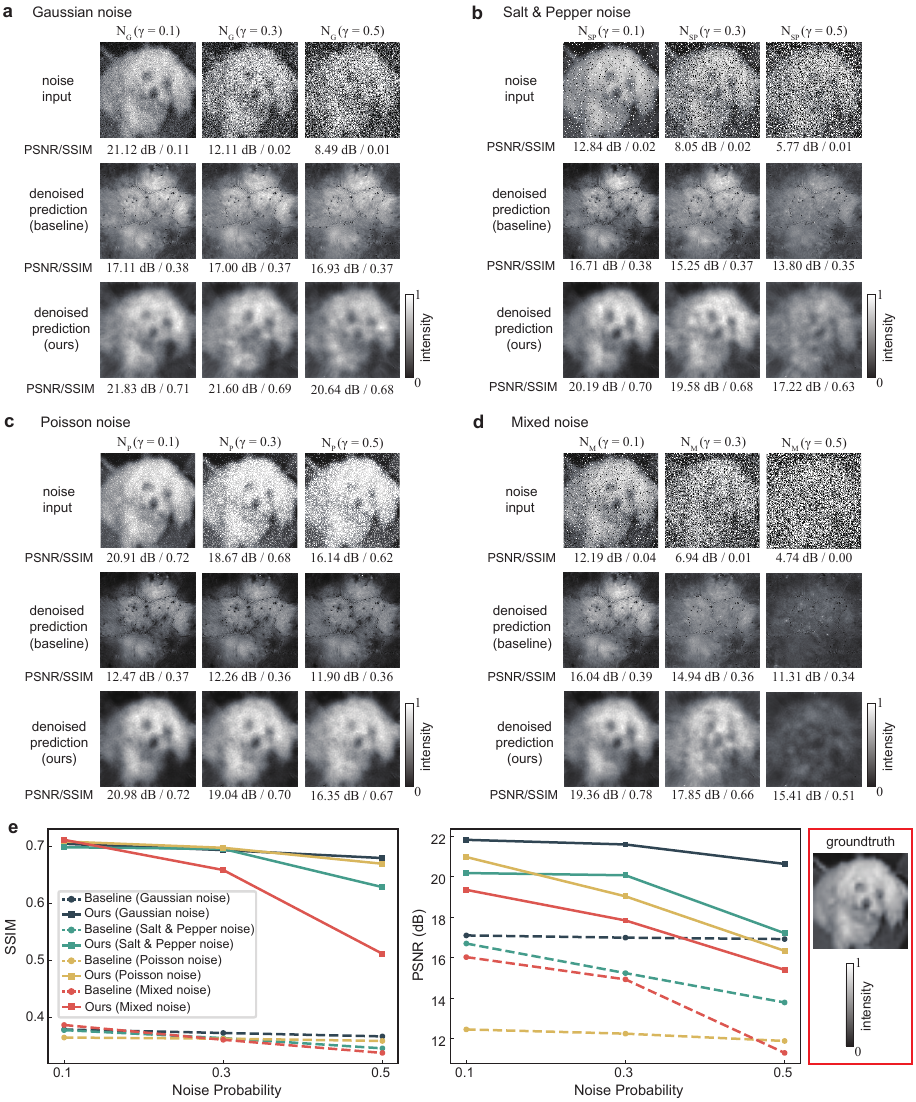}
    \caption{\textbf{Comparative analysis of baseline model (direct training) and ours model (pre-training and fine tuning) with different noise probability ($\gamma =\{0.1,0.3,0.5\}$).} The comparison results of the denoise performance with (a) Gaussian noise, (b) Salt \& Pepper noise, (c) Poisson noise, and (d) Mixer noise. (e) Quantitative assessment of denoising performance on the CIFAR-10 dataset, showing SSIM and PSNR metrics aggregated over 5000 test images.}
    \label{S10}
\end{figure}
The robustness of our diffractive denoising model was systematically evaluated against a baseline model across various noise types—including Gaussian $N_{G}$, Salt \& Pepper $N_{SP}$, Poisson $N_{P}$, and Mixed noise $N_{M}$—at varying levels of intensity (Figure \ref{S10}). Visual analysis of the CIFAR-10 dataset demonstrated that as the noise probability $\gamma$ increases from 0.1 to 0.5, the baseline model’s reconstruction quality degrades precipitously, frequently culminating in severe structural distortion or the complete eradication of critical semantic features. Contrarily, our optimized diffractive processor consistently maintained high-fidelity restorations, preserving sharp boundary definitions and intricate textural details even under high-noise regimes. This qualitative superiority was rooted in the model's learned capacity for non-local wavefront modulation, which effectively decoupled structured signals from stochastic interference. Such resilience was further substantiated by the quantitative trajectories of PSNR and SSIM metrics, where our approach exhibited a significantly more graceful performance decay than the baseline as noise levels escalate. Specifically, under the most demanding ``Mixed noise" conditions—which simulate complex real-world optical fluctuations by blending multiple stochastic sources—our diffractive processor retained substantial structural integrity and superior PSNR, underscoring its exceptional generalization capabilities and its potential as a robust front-end for processing non-ideal, high-dynamic optical signals.

\section{Comparison of structural parameters}\label{secA5}
\begin{figure}[thpb]
    \centering
    \includegraphics[width=\linewidth]{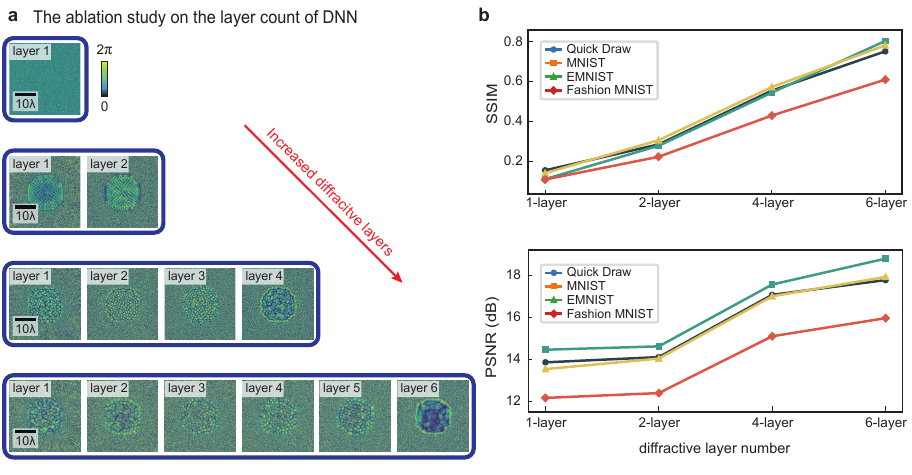}
    \caption{\textbf{Impact of the number of diffractive layers on denoising performance and phase mask evolution.} (a) Visualization of the optimized phase modulation patterns for diffractive networks with varying layer counts ($L$ = 1 to 6). (b) Quantitative comparison of denoising performance with different layer count, showing SSIM and PSNR metrics on four datasets.}
    \label{S11}
\end{figure}

\begin{figure}[thpb]
    \centering
    \includegraphics[width=\linewidth]{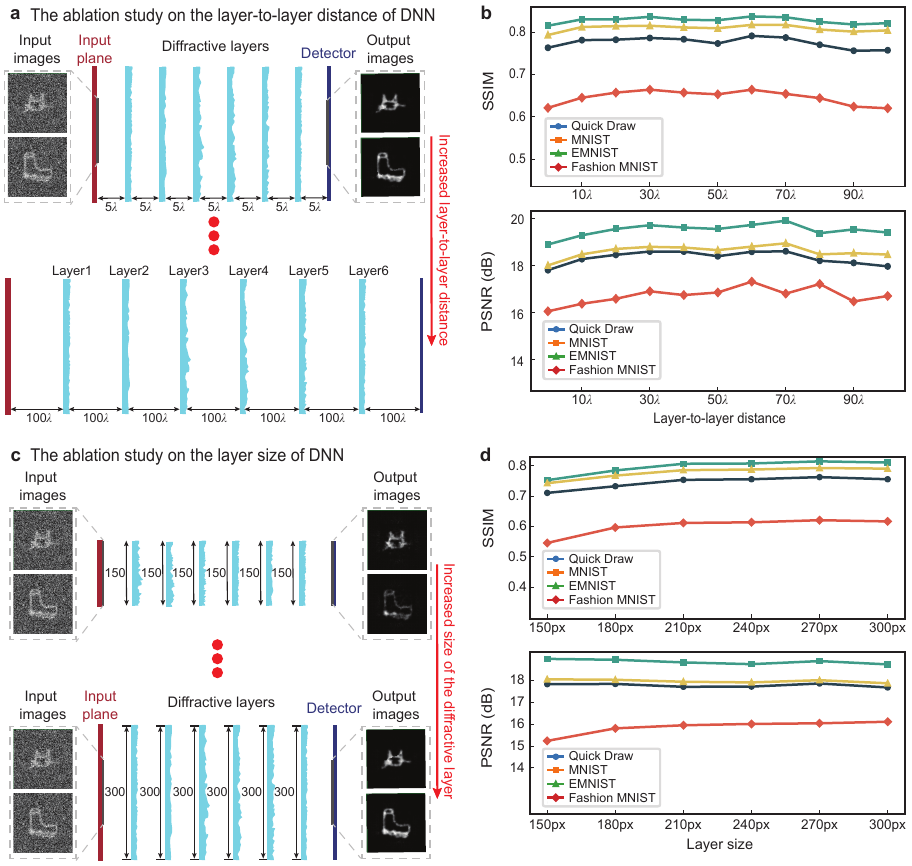}
    \caption{\textbf{Impact of the diffractive layer-to-layer distance and layer size on denoising performance.} (a) Visualization of the diffractive networks with varying layer-to-layer distance (from $5\lambda$ to $100\lambda$). (b) Quantitative comparison of denoising performance with different layer-to-layer distance, showing SSIM and PSNR metrics on four datasets. (c) Visualization of the diffractive networks with varying layer size (from $150px$ to $300px$). (d) Quantitative comparison of denoising performance with different layer size, showing SSIM and PSNR metrics on four datasets.}
    \label{S12}
\end{figure}

The ablation studies systematically evaluated the impact of key structural parameters—including the number of diffractive layers (Figure \ref{S11}), layer-to-layer distance, and layer size (Figure \ref{S12})—on the denoising performance of the all-optical diffractive processor. As the number of diffractive layers increased from 1 to 6, both the PSNR and SSIM demonstrated a consistent upward trend, suggesting that a higher depth in the optical network enhanced the capacity for complex wavefront modulation and noise suppression. Furthermore, the performance was sensitive to the distance between layers. An optimal layer-to-layer distance was observed where the diffractive features are most effectively propagated and interference patterns were best utilized for signal restoration. Similarly, increasing the physical size of the diffractive layers (number of modulation units) provided a larger computational area, which correlated with improved reconstruction fidelity, particularly for high-resolution semantic details. These parametric evaluations collectively identified an optimized structural configuration that balanced the device's physical footprint with its high-fidelity image restoration capabilities.

\section{Pre-training-enabled all-optical denoising for image classification}\label{secA6}
\begin{figure}[thpb]
    \centering
    \includegraphics[width=0.99\linewidth]{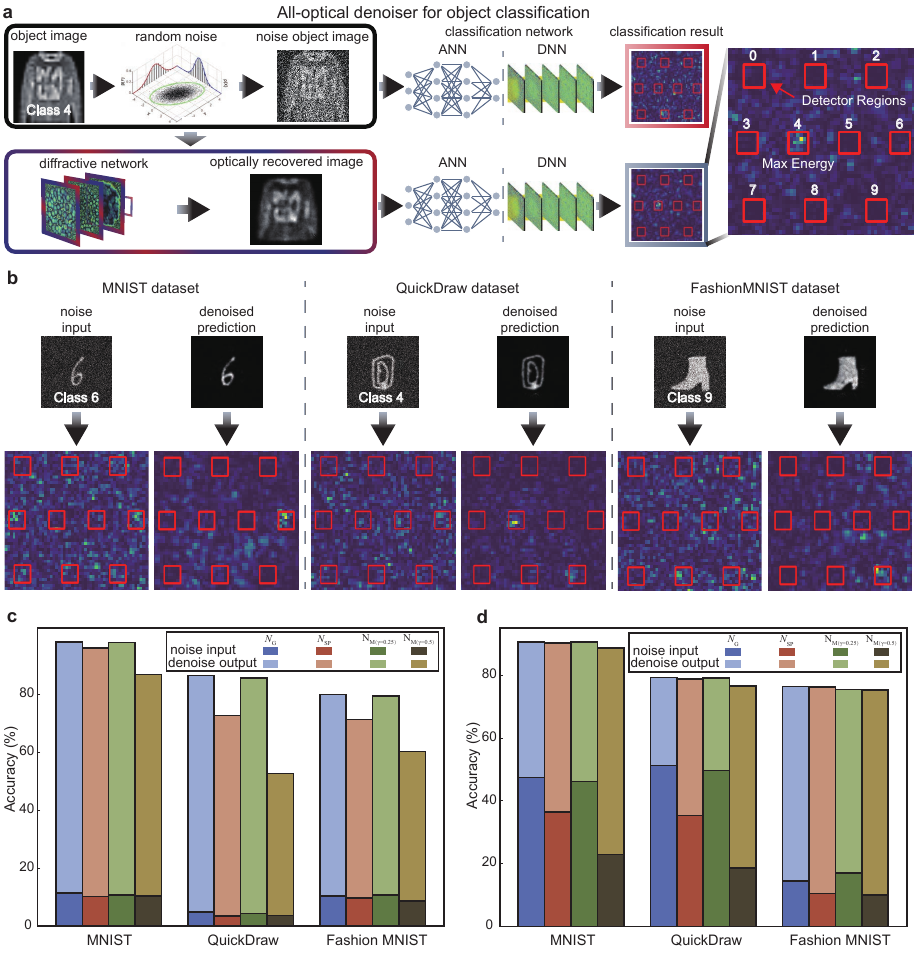}
    \caption{\textbf{Robustness enhancement of intelligent classification tasks via all-optical diffractive denoiser.} (a) Schematic pipeline of the hybrid optoelectronic vision system, where a 6-layer diffractive denoiser serves as a non-iterative front-end to restore noisy inputs before they are processed by either an optical neural network (ONN) or a digital artificial neural network (ANN) classifier. (b) Comparison of the optical power distribution and classification performance of the ONN. Quantitative evaluation of classification accuracy for the (c) ANN and (d) ONN  classifiers across various benchmark and complex datasets.}
    \label{S13}
\end{figure}
The integrated evaluation of our diffractive denoiser within a noise-adversarial classification framework demonstrated its significant utility as a high-speed, all-optical front-end for intelligent perception tasks (Figure \ref{S13}). The operational pipeline first employed the diffractive neural network to restore noisy input images before they were processed by either an optical neural network (ONN) or a digital artificial neural network (ANN) classifier. Visual and energy-distribution analysis of the ONN revealed that while stochastic noise severely dispersed the optical power at the detector plane—leading to ambiguous or incorrect classification—the introduction of the denoiser effectively re-concentrates the signal into the target detector regions, restoring sharp, high-contrast focal spots. Quantitative results further substantiated this across both analog and digital platforms. Both the ANN and ONN exhibited substantial improvements in classification accuracy when using denoised inputs compared to their raw, noisy counterparts across multiple benchmark datasets. This synergistic architecture highlighted the capability of the diffractive processor to mitigate the performance degradation caused by environmental interference, thereby enhancing the overall robustness and reliability of hybrid optoelectronic vision systems.